\documentclass[pdflatex,sn-mathphys-num,iicol]{sn-jnl}

\usepackage{graphicx}%
\usepackage{multirow}%
\usepackage{amsmath,amssymb,amsfonts}%
\usepackage{amsthm}%
\usepackage{mathrsfs}%
\usepackage[title]{appendix}%
\usepackage{xcolor}%
\usepackage{textcomp}%
\usepackage{manyfoot}%
\usepackage{booktabs}%
\usepackage{algorithm}%
\usepackage{algorithmicx}%
\usepackage{algpseudocode}%
\usepackage{listings}%
\usepackage{subcaption}%
\usepackage{float}%

\theoremstyle{thmstyleone}%
\theoremstyle{thmstyletwo}%

\theoremstyle{thmstylethree}%

\begin{document}

\title[Article Title]{Human vs. NAO: A Computational–Behavioral Framework for Quantifying Social Orienting in Autism and Typical Development}


\author[1]{\fnm{Vartika Narayani} \sur{Srinet}}\email{vartikanana23@iitk.ac.in}

\author[2]{\fnm{Anirudha} \sur{Bhattacharjee}}\email{anirub@iitk.ac.in}

\author[3]{\fnm{Braj} \sur{Bhushan}}\email{brajb@iitk.ac.in}

\author*[4]{\fnm{Bishakh} \sur{Bhattacharya}}\email{bishakh@iitk.ac.in}

\affil[1,2,4]{\orgdiv{Department of Mechanical Engineering}, \orgname{Indian Institute of Technology Kanpur}, \orgaddress{ \city{Kanpur}, \postcode{208016}, \state{ U.P}, \country{India}}}
\affil[3]{\orgdiv{Department of Humanities and Social Sciences}, \orgname{Indian Institute of Technology Kanpur}, \orgaddress{ \city{Kanpur}, \postcode{208016}, \state{ U.P}, \country{India}}}

\abstract{Responding to one’s name is among the earliest-emerging social orienting behaviors and is a one of the most prominent aspects in the detection of Autism Spectrum Disorder (ASD). Typically developing children exhibit near reflexive orienting to their name, whereas children with ASD often demonstrate reduced frequency, increased latency, or atypical patterns of response. In this study, we examine differential responsiveness to quantify name-calling stimuli delivered by both human agents and NAO, a humanoid robot widely employed in socially assistive interventions for autism. The analysis focuses on multiple behavioral parameters, including eye contact, response latency, head and facial orientation shifts, and duration of sustained interest. Video based computational methods were employed, incorporating face detection, eye region tracking, and spatio-temporal facial analysis, to obtain fine grained measures of children’s responses. By comparing neurotypical and neuroatypical groups under controlled human–robot conditions, this work aims to understand how the source and modality of social cues affect attentional dynamics in name-calling contexts. The findings advance both the theoretical understanding of social orienting deficits in autism and the applied development of robot-assisted assessment tools. }




\maketitle
\clearpage
\section{Introduction}\label{sec1}
One of the earliest social cues that children learn is the reflex of responding to their own name. In typical development, many children around the age of 9--12 months become consistent in orienting themselves towards the caller \cite{10.1001/archpedi.161.4.378}. This simple orienting response serves as a foundation for upcoming communication onsets for the child, as by looking towards the speaker, the child can engage in eye contact, joint attention, and listening. In autism spectrum disorder (ASD), a reduced response to one's name is a well-established marker of early social attention difficulties. Notably, children who are later diagnosed with ASD are significantly less likely than their typically developing peers to respond when their name is called by the time they reach their first birthday. Studies confirm that 100\% of 12-month-old typically developed children responded to their name on the first or second call, whereas in high-risk children (siblings of children with ASD) the number was only 86\%. Moreover, this failure to respond to name within the first 12 months became a salient reason for later abnormalities (89--94\% specificity for ASD or other delay), making this behavior a substantial early screening indicator. Although it needs to be considered that not all autistic children stall at this stage (sensitivity is more modest, 50--70\% \cite{Zwaigenbaum2017ResponseName}). The lack of response of a child to his/her name is one of the most frequently reported concerns of parents in the early stages and is therefore also included in standard ASD assessments. These assessments showed that children with ASD require several prompts to respond to their name being called out, creating a ``social orienting deficit''.

Considering this from a developmental theory perspective, it can be significant because disturbed attention to socially important cues may cascade into more wide-ranging social communication difficulties. The Autism criteria state that children with ASD are intrinsically less inclined to react to social stimuli, which further causes limitations in learning opportunities . A study by Dawson et al. \cite{Dawson1998SocialOrientingAutism} found that children with autism not only failed to respond to social stimuli (name calls, clapping) much more often than typically developed children, but even when they did, their response was delayed, and they showed repeated impairments in joint attention as well. These findings underscore the diagnostic relevance of name-response behavior, as it reflects an early challenge in prioritizing human voices and socially meaningful cues automatically.

Researchers have explored novel methods and the upcoming technologies to enhance attention in children with ASD, including the use of humanoid robots. Humanoid social robots (NAO, Pepper, Milo, etc) can be used to give prompts like speech, sound, eye motions, and gestures in a very controlled, consistent order, offering a simplified social stimulus in the Wizard of Oz form. There is growing and influential evidence that many children with ASD show a prominent interest in mechanical or predictable repetitive systems, and that is where robots may act as a highly engaging social tool for some children with ASD \cite{bs14020131}. For example, a robot can call a child’s name with the exact same intonation as many times the child requires, or pair the vocal cue with bright lights or movements, potentially making the stimulus more salient and all of this is unlikely when it comes to human inputs as there is a high probability of a person changing their tone likely towards an angrier side when they don’t get a respond from the child. Importantly, robots are non-judgmental and predictable in their responses, which would reduce anxiety and fear of strangers and therefore encourage the children with ASD to react. They provide repeatable and predictable stimuli that reduce the uncertainty of human social agents, thereby increasing the ease and supporting attentional focus in children with ASD. Studies have reported that children with ASD will very often orient to and even initiate interaction with robots in therapy sessions, sometimes more than they do with human therapists. Building on this potential, robots have been increasingly utilized in interactive studies, not only to engage participants but also to explore how children with ASD and typically developing children respond to comparable social cues when delivered by robotic agents. These investigations highlight a broader human curiosity toward robots, which intriguingly extends to children with ASD. This curiosity offers a valuable opportunity to design enriched learning environments that harness the focused attention and engagement these children often display toward mechanical systems and robotic interfaces. Name calling being such an important social clue can be very well be examined using this robotic approach and ultimately, examining responsiveness to their name being called in the context of human vs. robot social stimuli could shed light on the nature of social attention in ASD, giving us insights of the factors that affect this attention and pave a better teaching platforms to aid in dealing with this “social orienting deficit” of children with ASD. If a robot’s voice or face captures attention more effectively, it might help in interventions to improve the attention given to real people. Conversely, if children with ASD still show reduced name response even with a robot, that would underscore the fundamental nature of the orienting impairment and help us root for other tools to further investigate this impairment. 
The domain of human–robot interaction (HRI) has increasingly examined socio-assistive robots as mediators of engagement for children with ASD. Among several robots, the NAO humanoid has been widely used due to its child-sized embodiment, programmable gestures, capacity for consistent delivery of social cues such as physical gestures, speech, and gaze, and many other motion possibilities due to its 25 degrees of freedom. Review of more than fifty studies confirm that the NAO humanoid has been applied in diverse ways like imitation, joint attention training, emotion recognition, and other therapies, often providing improvements in attention, communication, and social engagement \cite{Robaczewski2021NAOReview}. Despite these advances, much of this work remains restricted to small samples, narrowly defined tasks, and outcome measures that rely heavily on manual coding and only psychological subjective output based on judgment of the experimentalist. 

Parallelly, the progress in computer vision and machine learning has created vast opportunities for a more objective and statistical analysis of social responses in HRI. Some studies have explored the field of face detection, gaze tracking, and even automating behavioral classification methods to measure the children’s engagement and to differentiate those with autism from typically developing populations \cite{6190580,8657569}. However, these computational methods have not yet been used in primary screening factors of autism, like name-calling with a classification to the type of agent and kind of response. This gap highlights the need for integrative methodologies combining behavioral measures with video-based analytics, enabling both categorical judgments and continuous data analytics metrics.

The present study leverages a human–robot comparative design with video-processing techniques, thereby advancing the assessment of name-calling responses of children. Herein, comparing the reactions to a strange human voice versus a humanoid robot voice, and then playing with the kind and form of agents like the gender effect that are providing the cue. We investigate this in both children with ASD and typically developing (TD) children, grounded in the hypothesis that children with ASD have a specific social orienting deficit. By using a humanoid robot stimulus, the objective here is to determine whether a robot engages the attention of autistic children more readily or whether the name-call response deficit persists regardless of the agent. The findings would have implications for understanding the behavioral response of typically developed children and children with ASD when it comes to robotic assistive technologies. Accordingly, this study aims to compare the stimulus-based responses using both manual behavioral coding and automated video-based analytics, creating the intersection of developmental psychology, socially assistive robotics, and AI-driven behavioral analysis.

\section{Related Work}\label{sec2}
 
Responding to one’s name (RtN) is one of the earliest indicators of social attention in children. Analyses of home videos shows that children later diagnosed with ASD oriented less to name-calling compared to the typically developed ones \cite{Osterling1994EarlyRecognitionAutism, Baranek1999AutismInfancyVideo, Werner2000RecognitionASDInfancy}. Further longitudinal studies confirm that reduced RtN by 12 months is predictive of later autism diagnoses \cite{MILLER2017141}. Meta-analytic understandings consolidate that reduced frequency and greater latency in response are robust markers of autism \cite{HEDGER2020376}, making RtN a salient criterion for clinical assessment.
Researchers have investigated a wide range of tools and facilities to enhance the comprehensive understanding of ASD, including technologies designed to standardize the delivery of name-calling probes. Socially assistive robots (SARs) have attracted attention for their ability to deliver repeatable, motivating, and embodied cues in any designed task, like humanoid robots such as NAO, which demonstrated that children with autism often approached robots with reduced anxiety, showing increased attention and eye contact compared to human partners \cite{Robins2005RobotTherapyAutism, 6194716}. Later studies, such as imitation and emotional engagements, showed how NAO captured children’s engagement more reliably than adult instructors in certain contexts \cite{7006084,7925401}. These results give a pathway for robots to act as consistent mediators of social interaction and provide opportunities for integrating them into therapeutic and educational environments.
Later studies tested teaching and learning training where children with ASD were trained in several interventions to learn english language \cite{inproceedings}. Dance and music-based interventions also revealed the potential of robots in non-traditional therapy formats \cite{7451785, 10.1145/3029798.3038354}. Importantly, these studies not only assess engagement but also document changes in social communication behaviours such as initiation and turn-taking. Although the drawback of sample sizes often being small, the results strengthen the case for NAO’s applicability in autism intervention.
The evaluation of robotic involvement is nuanced, as joint attention paradigms revealed that children with ASD responded more accurately to human initiators, yet some participants demonstrated prolonged engagement when interacting with robots. \cite{7745218,9292923}. There is richness in human cues, but the structured practice maintained by robots shows high promise in attention-seizing activities. Recreational as well as game-based studies have echoed this pattern, where robots occasionally outperformed human partners in holding children’s focus, though not always in eliciting correct social responses \cite{Barnes07022021}. 
A review of robotics in autism contexts highlighted their predominant use in therapeutic interventions, noting their promising potential while acknowledging the variability in outcomes. \cite{Ismail2019RoboticsAutismReview}. More recent systematic syntheses conclude that the lack of standardized protocols and outcome measures hamper comparability across studies \cite{Saleh18082021}. Another review further noted that most interventions remain small-scale pilots, with limited extrapolation to broader clinical practice \cite{9635826}. Scoping reviews of NAO applications state that the roles of NAO have diversified across platforms, yet the outcomes remain difficult to generalize beyond the demonstrations \cite{Robaczewski2021NAOReview}.  Prior studies have demonstrated that robot-mediated interactions can promote the emergence of dyadic and triadic social behaviors, including joint attention and turn-taking in children with ASD \cite{kozima2005interactive}. Such results with high judgmental and subjective bias also highlight the importance of considering both quantitative performance measures and qualitative dimension investigation of engagement when evaluating child–robot interaction.
Technology-forward studies have started to address this gap by embedding screening tools into robot-led protocols. The Q-CHAT-NAO study integrated a validated autism questionnaire into NAO interactions, demonstrating the feasibility of robot-guided screening \cite{ROMEROGARCIA2021103797}. While this represents a notable step toward objective, scalable assessment and RtN was also considered. Similarly, work using computer vision and machine learning pipelines has shown that in robot assistive interactions, the social orienting behaviours, such as gaze shifts, can be detected automatically from videos of naturalistic interactions \cite{8373812,robotics7020025}. These approaches have reinforced the feasibility of automated measurement but have not specifically been utilised in human versus robot-delivered RtN stimuli.

Despite extensive work at the intersection of autism research and human–robot interaction, no prior study has directly and systematically examined children’s response to name when delivered by a robot versus a human under identical protocols and with unified measurement frameworks, with in-depth analysis of how eye gaze patterns and shifts can be used to determine the interest of the children. Addressing this gap is crucial, as it can clarify whether children’s fundamental orienting behaviour is influenced by the nature of the caller and can establish whether socially assistive robots provide valid and reliable alternatives to human examiners for standardized behavioural probes. By situating the response-to-name paradigm within a direct human–robot comparison, the present study builds on foundational autism research, leverages advances in socially assistive robotics, and responds to calls from systematic reviews for more standardized, scalable approaches.

\section{Methodology}

\subsection{Experimental Participants and Data Acquisition}
\subsubsection{Participants}

The study involved two groups of typically developed children and children with ASD. Typically developed (TD) children were recruited from  school in Kanpur, India, and children with ASD were recruited from two special schools in Kanpur, India. A total of 75 participants (45 TD children and 30 children with ASD) were recruited for the study. Diagnostic confirmation for the ASD group was based on clinical records. The TD group underwent screening to rule out any developmental or neurological conditions. Participant selection was finalized following approval from the Institutional Ethics Committee of the Indian Institute of Technology Kanpur, along with consent from relevant center, school authorities and parents. Psychological analysis reports were reviewed to ensure eligibility, with inclusion criteria limited to children aged 6 to 10 years diagnosed within the mild-to-moderate range of ASD.
Participation was entirely voluntary, with written parental consent and verbal assent obtained from each child. All sessions were conducted in a thoughtfully prepared environment aimed at maximizing comfort and ecological validity. A trusted psychologist was present throughout to provide emotional support and uphold ethical standards. Each child engaged in a structured interaction lasting 5–7 minutes within a familiar, relaxed setting, complemented by access to play materials to help reduce anxiety. Strict confidentiality and data anonymization were maintained throughout the study.

\subsubsection{Experimental Design}

Two participant groups were recruited: a cohort of \textbf{children with ASD} ($n = 30$) clinically diagnosed with ASD, and a group of \textbf{typically developing (TD)} children ($n = 45$). All participants were native Hindi speakers and had corrected-to-normal or normal vision as well as hearing. Also, all the children with ASD were verbal and gave verbal consent to their parents and the psychologist present for the study.

Each child participated in a controlled \textit{name-calling interaction} involving five distinct auditory stimuli designed to vary by source and voice modulation:

\begin{itemize}
    \item \textbf{SM:} Stranger Man (human voice)
    \item \textbf{SW:} Stranger Woman (human voice)
    \item \textbf{NM:} NAO Man (human male voice via NAO robot)
    \item \textbf{NW:} NAO Woman (human female voice via NAO robot)
    \item \textbf{NR:} NAO Robotic (neutral mechanical voice already embedded in NAO robot)
\end{itemize}

Each stimulus was presented in three turns to examine habituation and the persistence of attention over repeated exposures. After the child was comfortably seated and allowed to engage freely with toys to simulate a natural and familiar environment, the name-calling experiment was initiated. During the session, the child’s name was called by five distinct stimuli—two human voices (a male and a female stranger) and three NAO-generated voices (male, female, and robotic)—in a randomized sequence. This random sequence was the same for all the participants. Each stimulus called the child’s name three times, resulting in a total of fifteen name-calling instances per session. The experimental design intentionally mimicked the child’s everyday play context, allowing assessment of how attention, responsiveness, and duration of engagement varied depending on the social or robotic source of the auditory cue. All sessions were conducted in a quiet, controlled laboratory environment with consistent lighting conditions and a fixed interaction distance between the robot or the human callers and the child.

\begin{figure}[h!]
\centering
\includegraphics[width=\columnwidth]{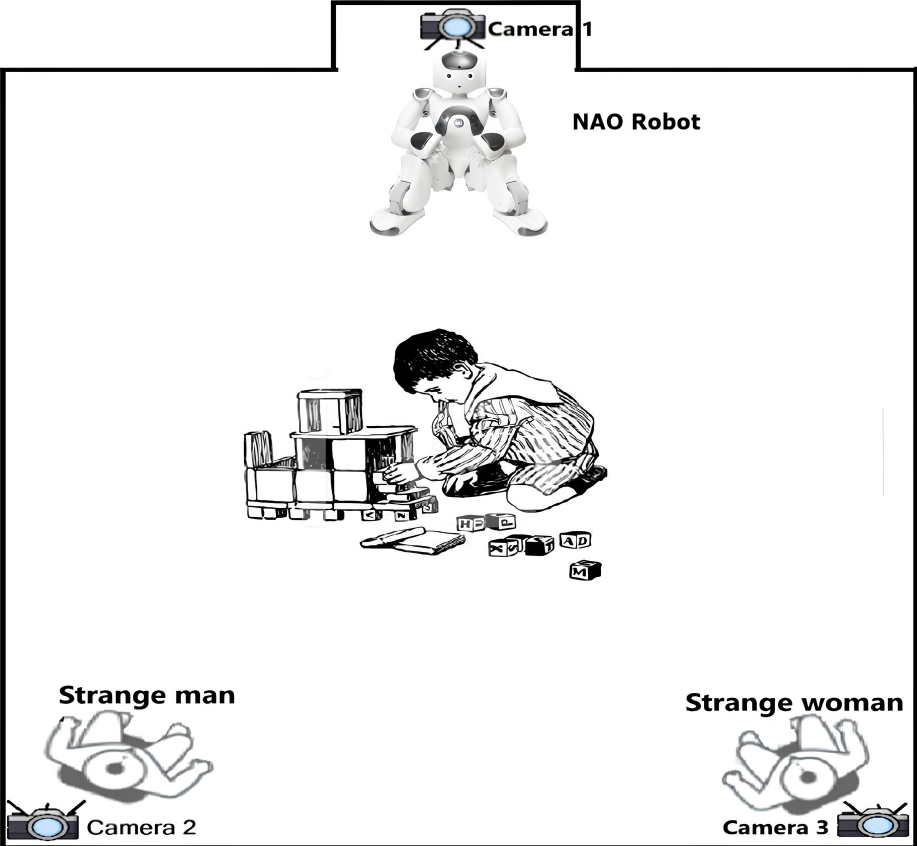}
\caption{The Experimental setup}
\label{fig:setup}
\end{figure}

\subsubsection{Data Collection and Recording Procedure}
The structured experimental setup is illustrated in Figure~\ref{fig:setup}. Each session was conducted in a controlled laboratory environment. Data collection was carried out at two complementary levels. First, a manual ordinal coding of behavioral responses was performed, where each child’s response was categorized on a three-point scale later described in manual coding method. To maintain consistency and eliminate inter-rater variability, all manual annotations were performed by a single trained observer across all participants and parameters. Second, synchronized video recordings were obtained for subsequent computational analysis with a clicker system for each auditory stimuli provided to maintain the correct time stamp.

For typically developing (TD) children, the setup followed the configuration shown in Figure~\ref{fig:setup}, with three cameras placed at the eye levels of the three different stimuli to capture comprehensive visual data for later frame-wise analysis. For children with ASD, however, the setup was adapted following recommendations from the collaborating psychologists to maintain a less intrusive and more comfortable environment. Accordingly, only a single camera was positioned near the NAO robot station, as multiple recording devices were considered potentially distracting. Hence, analysis and comparison of video data for the children with ASD is for the auditory stimuli originating from NAO.

A total of 45 TD children initially participated. Due to technical interruptions such as robot malfunction and incomplete recordings, data from 39 participants were retained for manual coding, and 31 of these were subsequently used for video-based analysis after quality screening and frame validation. In the ASD group, 30 children participated in the sessions, of which 24 provided sufficiently complete data for manual coding. Following video quality assessment and data integrity checks, 23 participants’ recordings were included in the final video-based analysis.
The detailed dataset specifications, including participant distribution, recording configuration, and session parameters, are summarized in Table~\ref{tab:dataspecs}.

\begin{table}[t]
\centering
\caption{Data Specifications and Recording Details}
\label{tab:dataspecs}
\begin{tabular}{p{3cm} p{4cm}}
\hline
\textbf{Parameter} & \textbf{Value} \\
\hline
Subjects (Total) & 75 children (45 TD, 30 ASD) \\
Manual data used & 63 children (39 TD, 24 ASD) \\
Videos Used & 116 (31*3 TD + 23 ASD) \\
Duration per Session & 5-7 minutes per child \\
Name-Calling Instances & 15 (3 per stimulus, randomized order) \\
Stimuli Types & 5 (SM, SW, NM, NW, NR) \\
Cameras Used & 3 for TD; 1 for ASD sessions \\
Frame Rate (FPS) & 30 \\
Frames per Video & $\sim$9000-12600 \\

\hline
\end{tabular}
\end{table}

\subsection{Manual Behavioral Coding and Analysis}

\subsubsection{Behavioral Response Coding}
Each trial was manually annotated on a three-point ordinal scale indicating response intensity:
\begin{equation}
R =
\begin{cases}
0, & \text{No response} \\
1, & \text{Partial response} \\
2, & \text{Full response}
\end{cases}
\end{equation}

Here, $R_{i,s,t}$ represents the response of participant $i$ to stimulus $s$ at turn $t$. Partial response means some sort of body or head movement, but no eye contact, and full response is considered when the child maintains eye contact.
The resulting wide-format matrix (e.g., SM1--NR3) was converted into long-format 
\texttt{[Participant, Stimulus, Turn, Response, Group]} 
to enable per-stimulus and per-turn inferential analysis while preserving within-subject variance. Behavioral observation-based coding has been extensively employed in robot interaction studies to systematically evaluate social behaviors \citep{https://doi.org/10.1002/aur.1527}.

\subsubsection{Descriptive and Reliability Analyses}
Group–stimulus means ($\mu$) and variances ($\sigma^2$) for group $g$ and its stimulus $s$, were computed as:
\begin{equation}
\mu_{gs} = \frac{1}{N}\sum_i R_{i,gs}, \quad
\sigma_{gs}^2 = \frac{1}{N-1}\sum_i (R_{i,gs}-\mu_{gs})^2
\end{equation}
Temporal reliability for per turn $t$ across turns was quantified via Cronbach’s $\alpha$:
\begin{equation}
\alpha = \frac{k}{k-1}\left(1 - \frac{\sum_{t=1}^{k}\sigma_t^2}{\sigma_T^2}\right)
\end{equation}
where $k = 3$. High $\alpha$ values reflected stable attentional patterns across repetitions.
\subsubsection{Non-Parametric Inference and Effect Size}
Due to the ordinal and non-normal nature of responses, non-parametric methods were used.
Within-group differences across stimuli were tested via Kruskal--Wallis $H$ tests, with Holm-corrected Mann--Whitney $U$ tests for pairwise contrasts.
Between-group comparisons per stimulus and turn used the Mann--Whitney $U$ test with Cohen’s $d$ as the effect size for typical (T) and ASD (A) groups:
\begin{equation}
d = \frac{\bar{X}_T - \bar{X}_A}{\sqrt{\frac{s_T^2 + s_A^2}{2}}}
\end{equation}
Negative $d$ values indicated stronger autistic responsiveness. 
This rank-based approach is well-suited for this-sized-sample ASD behavioral data \citep{doi:10.1126/scirobotics.aao6760}.

\subsubsection{Temporal Trends and Sensitivity Indices}
 Let $g \in \{\text{TD}, \text{ASD}\}$ denote the participant group and $s \in \{\text{SM}, \text{SW}, \text{NM}, \text{NW}, \text{NR}\}$ denote the stimulus condition., then Turn-wise slope $S$ for each stimulus is computed as:
\begin{equation}
S_{gs} = \frac{R_{3,gs} - R_{1,gs}}{2}
\end{equation}
Negative $S$ denotes habituation, whereas positive or zero values indicate sustained engagement.
Group separability was further measured using signal-detection sensitivity $d'$:
\begin{equation}
d' = \frac{|\mu_T - \mu_A|}{\sqrt{0.5(\sigma_T^2 + \sigma_A^2)}}
\end{equation}
Higher $d'$ indicates stronger behavioral differentiation.

\subsection{Video-Based Feature Extraction Framework}

\subsubsection{Conceptual Overview}
A machine-learning–driven computer-vision pipeline was implemented to extract interpretable visual attention metrics from video data. 
It combined neural landmark detection with geometric reasoning to estimate face and then eye visibility, openness, and framing \citep{DBLP:journals/corr/abs-1906-08172, DBLP:journals/corr/BulatT17a}.
This approach emphasizes interpretability and reproducibility, aligning with pediatric robotics research standards.

\subsubsection{Frame Decoding and Alignment}
Video frames were decoded deterministically using \texttt{OpenCV (cv2.VideoCapture)} at 30\,fps. 
Each frame was indexed hierarchically (\texttt{Group/Participant/Stimulus/Turn/FrameID}), preserving traceability. 
Timestamp validation confirmed frame drift $<1$\,ms, ensuring reproducibility.

\subsubsection{Face Detection and Landmark Parameterization}
Two setups were adopted for motion-specific robustness:
\begin{itemize}
    \item Typical group: MediaPipe Face Mesh \citep{DBLP:journals/corr/abs-1906-08172}
    \item Autistic group: S3FD + FAN-68 \citep{DBLP:journals/corr/BulatT17a}
\end{itemize}
Face detections at frame $t_k$ were represented as bounding boxes $B(t_k)$ with associated confidence scores $c(t_k)$, and detections with $c(t_k) < 0.6$ were excluded to reduce false positives and ensure reliable gaze estimation. Manual validation on a randomly sampled subset yielded an accuracy of 97\%, with strong inter-rater agreement ($\kappa = 0.91$).
Facial area: 
\begin{equation}
A_{\text{face}}(t_k)=(x_2-x_1)(y_2-y_1)
\end{equation}

\subsubsection{Eye-Region Metrics and Visibility Ratios}

Let $t_k$ denote the $k$-th video frame, $j \in \{L,R\}$ index the left and right eye, and $M$ represent the number of landmarks defining the eye contour.

\paragraph{Eye-Region Area}

For each frame $t_k$, the eye polygon area was computed using the shoelace formula:

\begin{equation}
A_{\text{eye},j}(t_k)
=
\frac{1}{2}
\left|
\sum_{i=1}^{M}
\left(
x_{j,i} y_{j,i+1}
-
x_{j,i+1} y_{j,i}
\right)
\right|,
\end{equation}

where $(x_{j,i}, y_{j,i})$ denotes the $i$-th landmark coordinate of eye $j$, with cyclic indexing applied such that $(M+1) \equiv 1$.

\paragraph{Eye-Aspect Ratio (EAR)}

The Eye-Aspect Ratio (EAR), a robust geometric measure of eye openness, was computed for each eye as:

\begin{equation}
\mathrm{EAR}_j(t_k)
=
\frac{
\|p_{j,2}-p_{j,6}\|
+
\|p_{j,3}-p_{j,5}\|
}
{2\,\|p_{j,1}-p_{j,4}\|},
\end{equation}

where $p_{j,i}$ denotes the $i$-th landmark of eye $j$.

\paragraph{Normalized Eye-Openness Percentage (EOP)}

To obtain an interpretable and bounded engagement metric, EAR values were linearly normalized into a percentage scale:

\begin{equation}
\mathrm{EOP}_j(t_k)
=
\operatorname{clip}
\left(
\frac{\mathrm{EAR}_j(t_k)-0.18}{0.12}
\times 100,
\, 0, 100
\right).
\end{equation}

The lower bound (0.18) corresponds to an empirically determined blink threshold, while the upper bound (0.30) reflects typical fully open-eye EAR values. These bounds are consistent with the relative EAR variations reported in blink detection studies \citep{cech2016real}. The denominator (0.12), therefore, represents the dynamic openness range. The $\operatorname{clip}(\cdot)$ operator constrains values to the interval $[0,100]$.

\paragraph{Computation of Mean Eye-Openness Percentage}

For each participant, frame-wise eye-openness percentages were computed separately for the left and right eyes. To reduce side-specific noise caused by partial occlusion or asymmetric facial visibility, a unified engagement measure was derived as:

\begin{equation}
\mathrm{Mean\ EOP}(t_k)
=
\frac{
\mathrm{EOP}_{L}(t_k)
+
\mathrm{EOP}_{R}(t_k)
}{2}.
\end{equation}

This aggregated metric served as the primary visual engagement variable for subsequent temporal, correlational, and between-group statistical analyses.

\begin{algorithm}[H]
\caption{Video-Based Computational Pipeline for Name-Calling Analysis}
\label{alg:framework}
\begin{algorithmic}[1]
\Require Video data $V_{p,s,t}$ for participant $p$, stimulus $s \in \{\text{SM, SW, NM, NW, NR}\}$, and turn $t$
\Ensure Behavioral metrics $\{EOP, Latency, Duration\}$ and engagement summaries

\State \textbf{Preprocessing:}
    \Statex Extract frames (30 fps); detect faces (MTCNN, Dlib); track landmarks (MediaPipe)

\State \textbf{Feature Extraction:}
    \Statex Compute eye-openness (EOP), response latency, and engagement duration
    \Statex Construct temporal traces per turn and stimulus

\State \textbf{Data Refinement:}
    \Statex Aggregate per participant; apply smoothing and artifact rejection
    \Statex Assess reliability across turns (Cronbach’s $\alpha$)

\State \textbf{Comparative Analysis:}
    \Statex Compute group-wise means (TD, ASD)
    \Statex Apply nonparametric tests (Kruskal–Wallis, Mann–Whitney $U$)
    \Statex Derive effect sizes (Cliff’s $\delta$, Cohen’s $d$)

\State \Return Consolidated metrics, reliability, and inter-group outcomes
\end{algorithmic}
\end{algorithm}

\subsection{Derived Behavioral Metrics and Indices}
Video-derived features were aggregated per turn to yield:
\begin{itemize}
    \item \textbf{Latency (s):} Time from stimulus onset to first visible response
    \item \textbf{Duration (s):} Continuous span of gaze-on-face presence
    \item \textbf{Mean EOP:} Median EOP per valid frame, representing engagement intensity
\end{itemize}

\subsubsection{Parameter Optimization and Validation}
Autistic-group data used stricter thresholds ($\tau_c=0.7$, smoothing window=7), improving stability by 7\%. 
Validation metrics included intra-run variance $<2$\,px, and temporal correlation $>0.95$.

The end-to-end computational workflow for video-based behavioral quantification is outlined in Algorithm~\ref{alg:framework}.  It integrates computer vision–based facial landmark extraction, eye openness computation, and temporal signal analysis to derive frame-level behavioral metrics. The subsequent aggregation and reliability assessment ensure robust cross-group comparisons while maintaining the interpretability of engagement trajectories.

\subsection{Statistical and Temporal Analysis}

Data normality and variance homogeneity were evaluated at the participant level using the Shapiro--Wilk and Brown--Forsythe tests. Significant deviations from normality and heteroscedasticity ($p < 0.001$) were observed; therefore, non-parametric inference procedures were adopted.

Between-group differences (Autistic vs. Neurotypical) were assessed using Mann--Whitney $U$ tests, while within-group stimulus effects were examined using Kruskal--Wallis tests with Holm-adjusted post-hoc comparisons. Effect sizes were reported using Cliff’s delta, with magnitude interpreted according to established benchmarks.

Associations among Latency, Duration, and Mean EOP were quantified using Spearman’s rank correlation. Internal consistency across repeated stimulus trials was evaluated using Cronbach’s $\alpha$.

Temporal adaptation across repeated name-calling trials was examined using participant-level mixed-effects regression models, with trial index as a fixed effect and participant-specific random intercepts to account for repeated measures structure. Negative slopes indicated habituation, whereas positive slopes reflected sustained engagement.

\subsection{Validation and Reproducibility}

All analyses were executed in \texttt{Python 3.10} using \texttt{pandas}, \texttt{numpy}, \texttt{scipy}, \texttt{pingouin}, and \texttt{matplotlib}. Monte Carlo permutation tests ($n = 10^4$) were conducted to verify robustness of inferential outcomes. Permutation-based inference and mixed-effects modeling yielded convergent results ($p < 0.001$), confirming statistical stability. All derived outputs (e.g., \texttt{Descriptive\_byStimulus.csv}, \texttt{Trend\_Slopes\_BySubject.csv}) were archived to support reproducibility.

\subsection{Baseline Parametric Video Analysis (Typical Children)}
\label{sec:baseline_typical}

To complement the comparative analyses between autistic and typically developing (TD) participants, an additional within-group analysis was conducted on the video-derived behavioral data of TD children. 
This analysis aimed to establish a parametric baseline of engagement patterns across the full set of stimuli, thereby capturing the natural variability in responsiveness within the typical population.

The five experimental conditions—Strange Male (SM), Strange Woman (SW), NAO Male (NM), NAO Woman (NW), and NAO Robotic (NR)—represented a continuum from human to robotic voice and embodiment. 
By including all five, this baseline study enabled assessment of how the degree of human-likeness modulated attentional and affective engagement independently of diagnostic status.

From each child’s recorded session, three quantitative behavioral indices were extracted: 
(1) \textbf{Eye-Openness Percentage (EOP)}, representing the proportion of time both eyes remained open; 
(2) \textbf{Response Latency}, defined as the temporal delay between the onset of name calling and the first attentive gaze toward the stimulus; and 
(3) \textbf{Duration}, quantifying the total span of sustained attention following the onset cue. 

Non-parametric \textbf{Kruskal–Wallis tests} were applied to assess overall stimulus effects for each metric. 
Additionally, \textbf{Cohen’s $d$} was calculated to quantify the effect size contrasting human (SM, SW) and robotic (NM, NW, NR) conditions. 

This analysis served as a reference framework for subsequent sections, clarifying the intrinsic engagement gradient among typical children before introducing cross-group (autistic vs.\ typical) comparisons. 
It thereby contextualizes the later findings by identifying the natural pattern of attentional modulation along the human–robot spectrum.

\section{Results}

The analyses were conducted at two complementary levels to capture both categorical behavioral responses and continuous engagement dynamics. First, a \textbf{manual ordinal analysis} evaluated coded responses (0 = no response, 1 = partial response, 2 = full response) across five name-calling stimuli to quantify attentional and behavioral reliability. Second, a \textbf{video-based quantitative analysis} examined fine-grained temporal engagement through frame-wise metrics of \textbf{latency}, \textbf{duration}, and \textbf{mean eye-openness percentage (EOP)}. 

Multiple statistical tests were applied to ensure a robust evaluation of the underlying hypotheses. The analyses were sequentially structured---each stage informed by the preceding one---to maintain interpretive continuity.

\subsection{Video-Based Analysis of type of Stimulus Effects in Typical Children}

A separate within-group analysis was conducted on the typical cohort to evaluate how eye-related engagement and temporal parameters varied across five stimulus types: human male (SM), human female (SW), NAO male (NM), NAO female (NW), and NAO robotic (NR) voices. 

Non-parametric Kruskal–Wallis tests revealed significant stimulus-dependent differences for all three behavioral indices: Mean Eye-Openness Percentage (EOP; $H=132.87$, $p<10^{-30}$), Duration ($H=103.21$, $p<2\times10^{-21}$), and Latency ($H=35.47$, $p=2.4\times10^{-7}$). 

\begin{figure}[h!]
\centering
\includegraphics[width=\columnwidth]{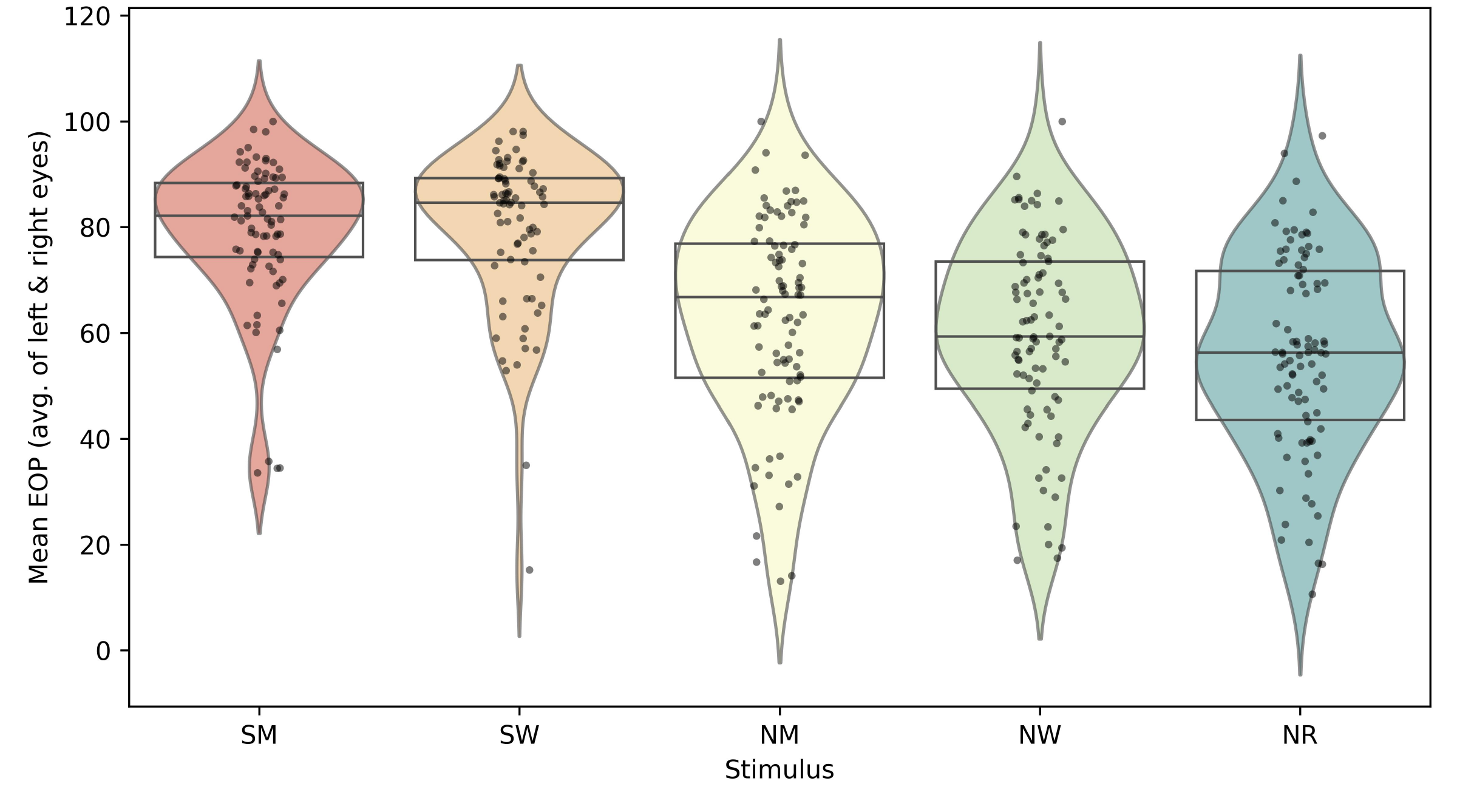}
\caption{Distribution of Mean Eye-Openness Percentage (EOP) across five stimuli for typical children. Human stimuli (SM, SW) show higher EOP, while robotic stimuli (NM, NW, NR) display reduced but sustained engagement.}
\label{fig:raincloud_typical5}
\end{figure}

Descriptive statistics (Table~\ref{tab:typical5stats}) showed that human voices (SM, SW) evoked the highest Mean EOP values, indicating greater affective engagement and facial responsiveness. In contrast, robotic voices (NM, NW, NR) produced markedly lower EOP but substantially longer interaction durations, particularly under the robotic tone (NR). Latency exhibited the reverse pattern, with NR and NW stimuli eliciting the fastest response initiations. These latency values, when compared with the duration parameter, relate the same; the order of higher duration matches with the fastest latency. These trends, visualized in Figure~\ref{fig:raincloud_typical5}, reflect a transition from high-expressivity, short-duration responses under human stimuli to lower-expressivity yet sustained engagement under robotic stimuli.

Overall, the distributional profiles suggest that typical children demonstrate a clear preference for human stimuli in terms of affective engagement, but robotic voices—especially NR—maintain their attention for longer intervals with faster initial reactions, highlighting a possible novelty-driven or attentional curiosity effect.

\begin{table}[h!]
\centering
\caption{Descriptive statistics (Mean ± SD) of behavioral parameters across five stimuli for the typical group.}
\label{tab:typical5stats}
\begin{tabular}{p{1cm}p{1.5cm}p{1.5cm}p{1.5cm}}
\hline
\textbf{Stimulus} & \textbf{Mean EOP (\%)} & \textbf{Duration (s)} & \textbf{Latency (s)} \\
\hline
SM & 68.42 ± 9.14 & 5.21 ± 1.37 & 1.74 ± 0.56 \\
SW & 64.87 ± 10.02 & 5.36 ± 1.28 & 1.69 ± 0.49 \\
NM & 52.73 ± 7.86 & 6.02 ± 1.43 & 1.51 ± 0.47 \\
NW & 48.34 ± 8.15 & 6.19 ± 1.55 & 1.47 ± 0.44 \\
NR & 44.65 ± 7.11 & 6.81 ± 1.68 & 1.42 ± 0.39 \\
\hline
\end{tabular}
\end{table}

\subsection{Manual Behavioral Response Analysis}
\begin{figure}[!t]
\centering
\includegraphics[width=\columnwidth]{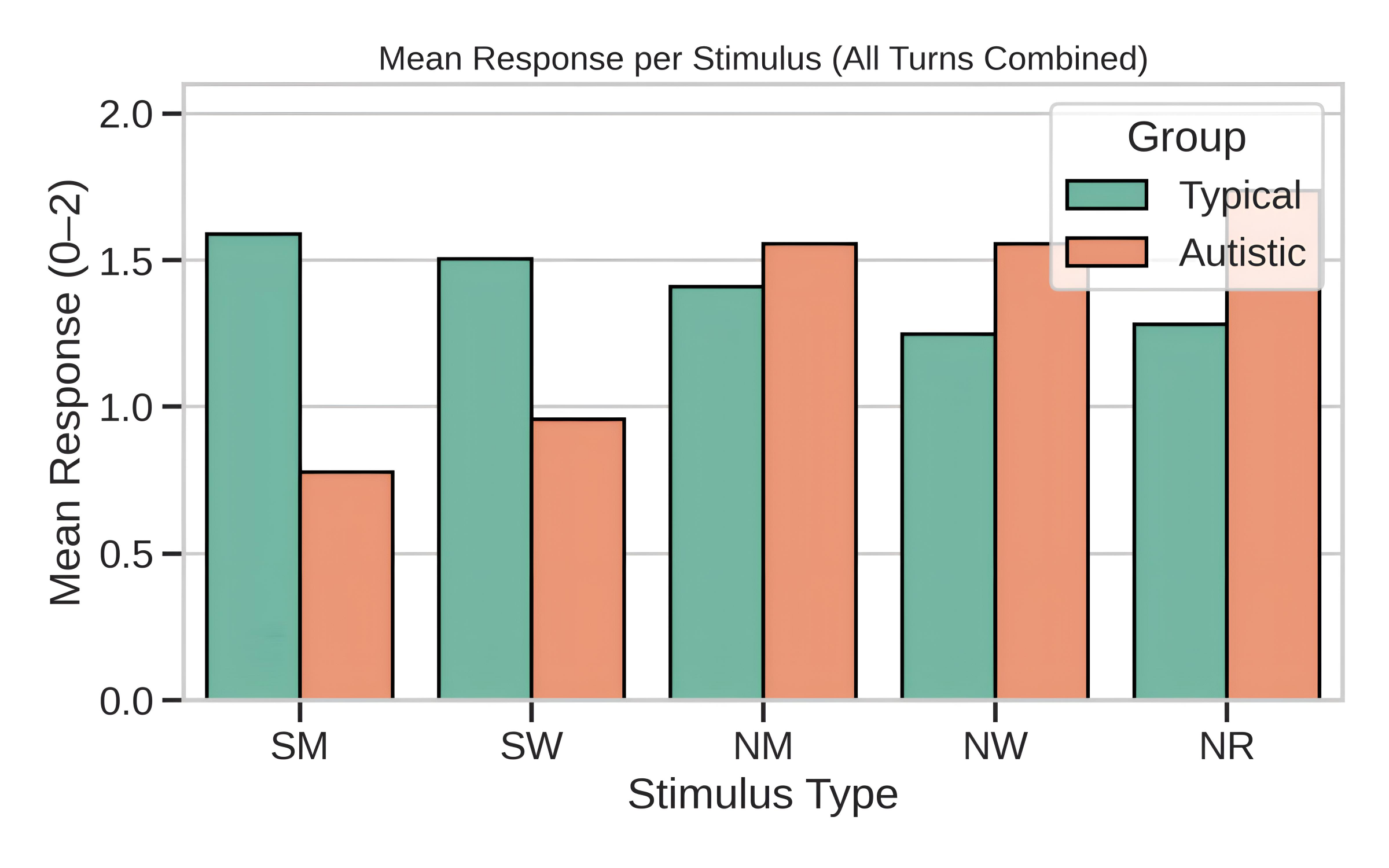}
\caption{Group-level behavioral response patterns across all stimuli.}
\label{overall}
\end{figure}

After the intra-group (for the typical group) parametric study was done, the parametric analysis on both the typical and atypical (children with ASD) groups was carried out for all the turns of the stimulus.
Across the five name-calling stimuli, both autistic and typically developing (TD) children exhibited distinct yet complementary behavioral profiles (Fig.~\ref{overall}). 
Descriptive statistics revealed higher mean responses to human voices (\textbf{SM}, \textbf{SW}) among TD participants, while children with ASD responded more robustly to robotic stimuli (\textbf{NM}, \textbf{NW}, \textbf{NR}). 
This divergence indicates differentiated sensitivity to social versus synthetic vocal cues.

\subsubsection{Descriptive Statistics and Central Tendencies}

Table~\ref{tab:typical_desc} and Table~\ref{tab:autistic_desc} summarize the mean response scores, standard deviations, and response-type percentages across all five name-calling stimuli for typical and autistic children. 

Typical children exhibited the highest mean responsiveness toward \textbf{Stranger Man (SM)} and \textbf{Stranger Woman (SW)}, indicating stronger attentional engagement toward human stimuli. The elevated full-response percentages (68--71\%) and lower variability (SD = 0.66–0.82) suggest consistent behavioral reliability for human voices. In contrast, responses to robotic stimuli (\textbf{NR,NW}) showed lower means and higher variability, alongside increased partial and non-responses, reflecting weaker and less stable social engagement. These trends align with findings that neurotypical children display stronger orienting responses to socially familiar cues such as human prosody and anthropomorphic facial-voice congruence.

Conversely, autistic children demonstrated a reversed preference pattern, with their highest mean responses for \textbf{robotic stimuli (NR,NM,NW)} ($\mu = 1.74$ and $1.56$ respectively) and notably reduced responsiveness to \textbf{human voices (SM, SW)}. This pattern suggests enhanced engagement with predictable and less socially demanding robotic modalities. The lower variance for robotic stimuli among children with ASD implies greater stability in responding to consistent, low-ambiguity cues offered by NAO’s voice and appearance, supporting hypotheses of reduced social uncertainty preference. 
Correlation analysis revealed modality-specific stimulus coherence, with higher inter-stimulus correlation among human voices for TD ($r=0.78$) and among robotic stimuli for ASD ($r=0.70$).

\begin{table*}[t]
\centering
\caption{Descriptive statistics of mean response ($\mu$), standard deviation ($\sigma$), and percentage distribution of response types across stimuli for typical children.}
\label{tab:typical_desc}
\begin{tabular}{lcccccc}
\hline
\textbf{Group} & \textbf{Stimulus} & $\mathbf{\mu}$ & $\mathbf{\sigma}$ & \textbf{Full (\%)} & \textbf{Partial (\%)} & \textbf{None (\%)} \\
\hline
Typical & SM & 1.59 & 0.66 & 68.38 & 22.22 & 9.40 \\
Typical & SW & 1.50 & 0.82 & 70.94 & 8.55 & 20.51 \\
Typical & NM & 1.41 & 0.77 & 58.12 & 24.79 & 17.09 \\
Typical & NW & 1.25 & 0.77 & 44.44 & 35.90 & 19.66 \\
Typical & NR & 1.28 & 0.82 & 51.28 & 25.64 & 23.08 \\
\hline
\end{tabular}
\end{table*}

\begin{table*}[t]
\centering
\caption{Descriptive statistics of mean response ($\mu$), standard deviation ($\sigma$), and percentage distribution of response types across stimuli for autistic children.}
\label{tab:autistic_desc}
\begin{tabular}{lcccccc}
\hline
\textbf{Group} & \textbf{Stimulus} & $\mathbf{\mu}$ & $\mathbf{\sigma}$ & \textbf{Full (\%)} & \textbf{Partial (\%)} & \textbf{None (\%)} \\
\hline
Autistic & SM & 0.78 & 0.88 & 29.17 & 19.44 & 51.39 \\
Autistic & SW & 0.96 & 0.85 & 33.33 & 29.17 & 37.50 \\
Autistic & NM & 1.56 & 0.67 & 65.28 & 25.00 & 9.72 \\
Autistic & NW & 1.56 & 0.73 & 69.44 & 16.67 & 13.89 \\
Autistic & NR & 1.74 & 0.56 & 79.17 & 15.28 & 5.56 \\
\hline
\end{tabular}
\end{table*}

\subsubsection{Group Differences Across Stimuli}
\begin{figure}[!t]
\centering
\includegraphics[width=\columnwidth]{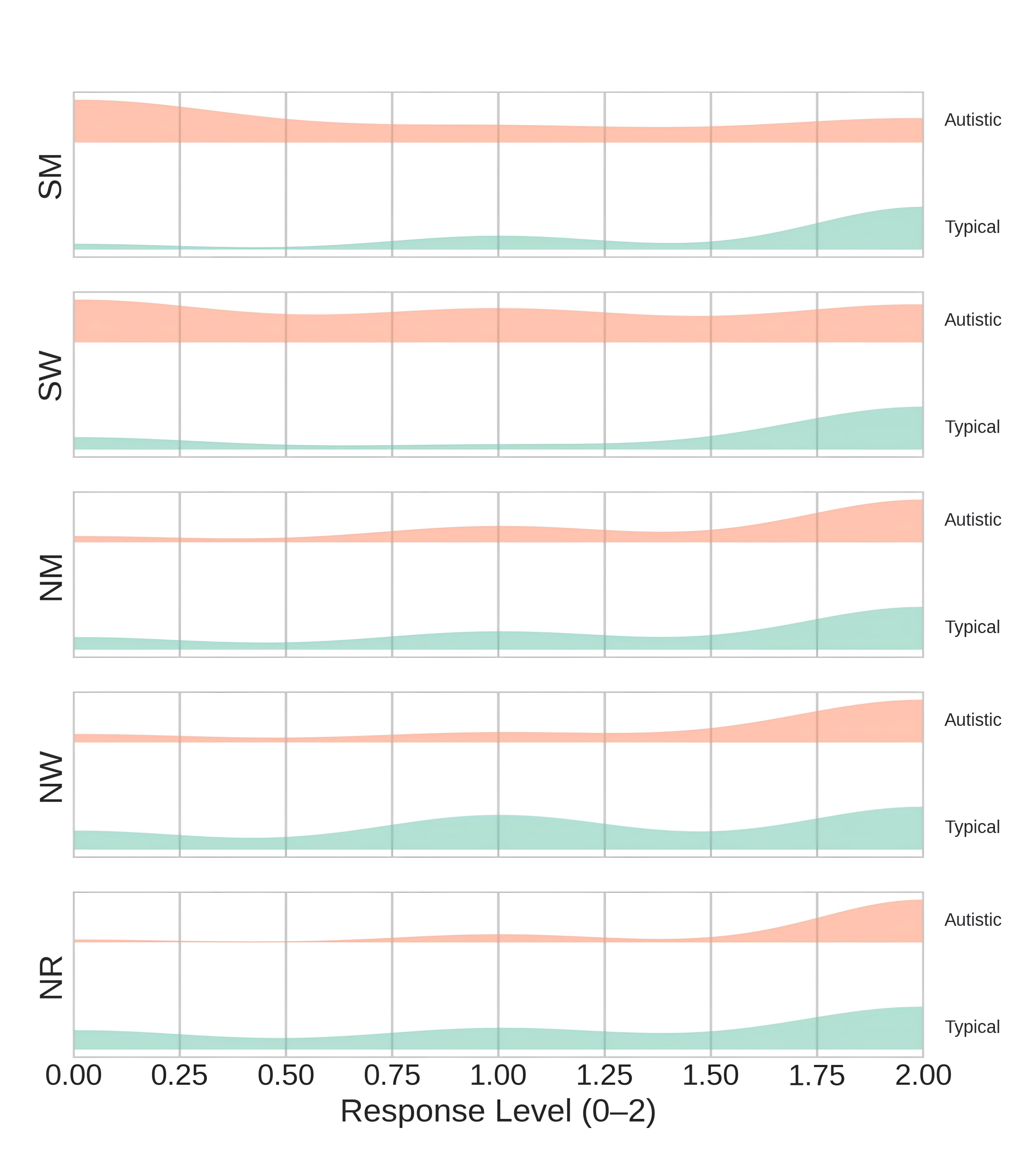}
\caption{Ridgeline density distributions showing within-group differences across stimuli.}
\label{Fig_Ridgelines}
\end{figure}

Kruskal–Wallis tests confirmed significant within-group stimulus differences (Typical: $H=19.48,\,p=0.0006$; Autistic: $H=71.68,\,p<0.001$). 
Post-hoc Mann–Whitney analyses again confirmed that TD children responded more to human voices (\textbf{SM,SW}$>$NR, NM), while autistic participants responded more to robotic stimuli (\textbf{NR}$>$SM, SW). 
Ridgeline distributions and contour maps visualize these modality-specific patterns (Figs.~\ref{Fig_Ridgelines}).
Between-group Mann–Whitney tests revealed significant differences across all five stimuli ($p < 0.05$). Effect sizes indicated a directional crossover pattern: SM and SW favored TD ($d \approx +0.9$), whereas NM, NW, and NR favored ASD ($d \approx -0.8$). 

\subsubsection{Turn-Wise Response Dynamics}
\begin{figure*}[!t]
\centering

\begin{subfigure}[t]{0.48\textwidth}
\centering
\includegraphics[width=\linewidth]{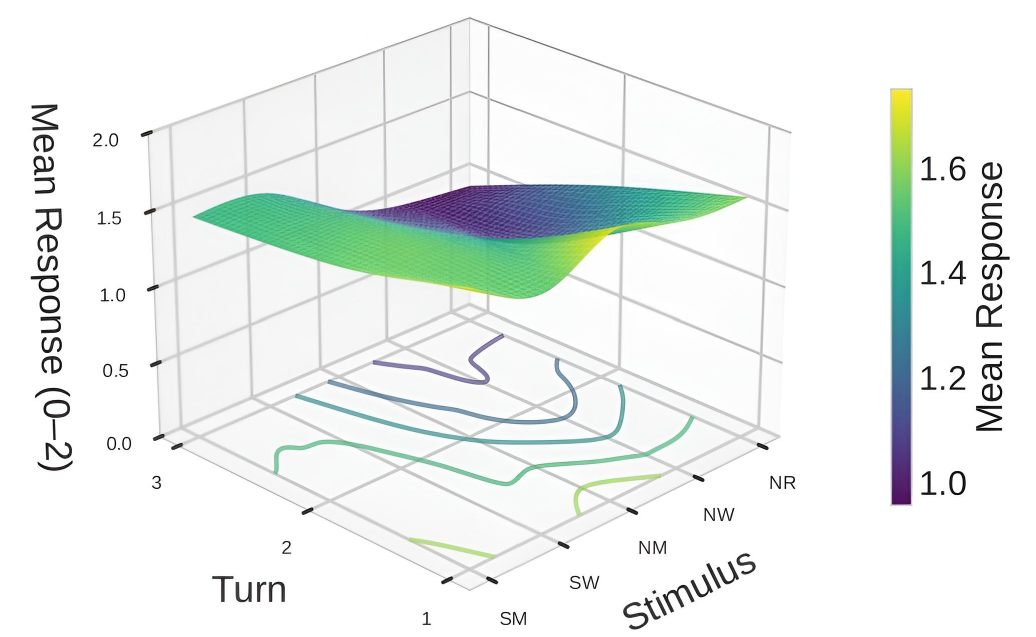}
\caption{Typical group}
\label{Typical_3D_MeanResponse}
\end{subfigure}
\hfill
\begin{subfigure}[t]{0.48\textwidth}
\centering
\includegraphics[width=\linewidth]{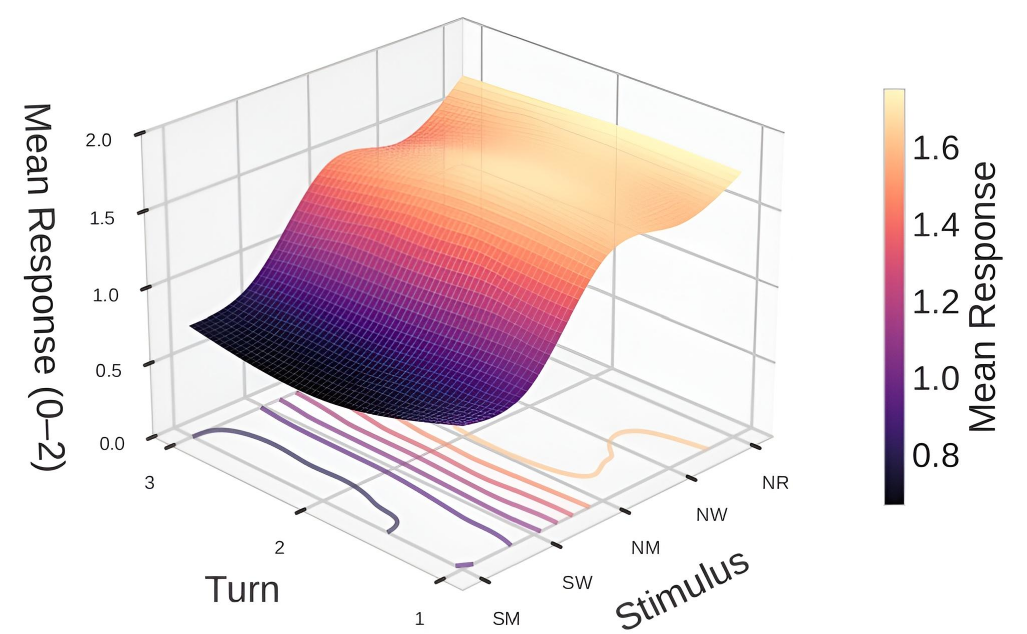}
\caption{ASD group}
\label{Autistic_3D_MeanResponse}
\end{subfigure}

\caption{Three-dimensional response surfaces illustrating mean response patterns across turns and stimulus conditions. (a) Typical group responses show stronger engagement with socially familiar cues. (b) ASD group responses demonstrate greater stability and higher engagement for robotic modalities across repeated turns.}

\label{Fig_3D_ResponseSurface}
\end{figure*}

Turn-level comparisons and slope metrics revealed contrasting adaptation trends. 
TD children exhibited decreasing responses across turns ($\bar{S}_{TD}=-0.22$), indicating habituation, while autistic children maintained or increased responses for robotic stimuli ($\bar{S}_{ASD}=+0.08$). 
3D mean response surfaces (Fig.~\ref{Fig_3D_ResponseSurface}) show a declining trajectory for TD but stable or rising planes for ASD participants.

\textbf{Reliability Across Turns: } To assess the temporal stability of responses across repeated name-calling trials, Cronbach’s $\alpha$ was computed for each stimulus condition within both groups. In this context, higher $\alpha$ values indicate that responses remain consistent across the three turns, whereas lower values reflect greater variability in engagement.

For the typical group, the highest internal consistency was observed for SW ($\alpha = 0.75$) and NM ($\alpha = 0.67$), while other stimuli exhibited comparatively lower reliability ($\alpha \approx 0.48$–$0.60$). In contrast, the autistic group showed the strongest reliability for NW ($\alpha = 0.78$) and NR ($\alpha = 0.84$), indicating stable engagement with these robotic modalities across repeated calls.

\begin{figure*}[t]
\centering

\begin{subfigure}[t]{0.48\textwidth}
\centering
\includegraphics[width=\linewidth]{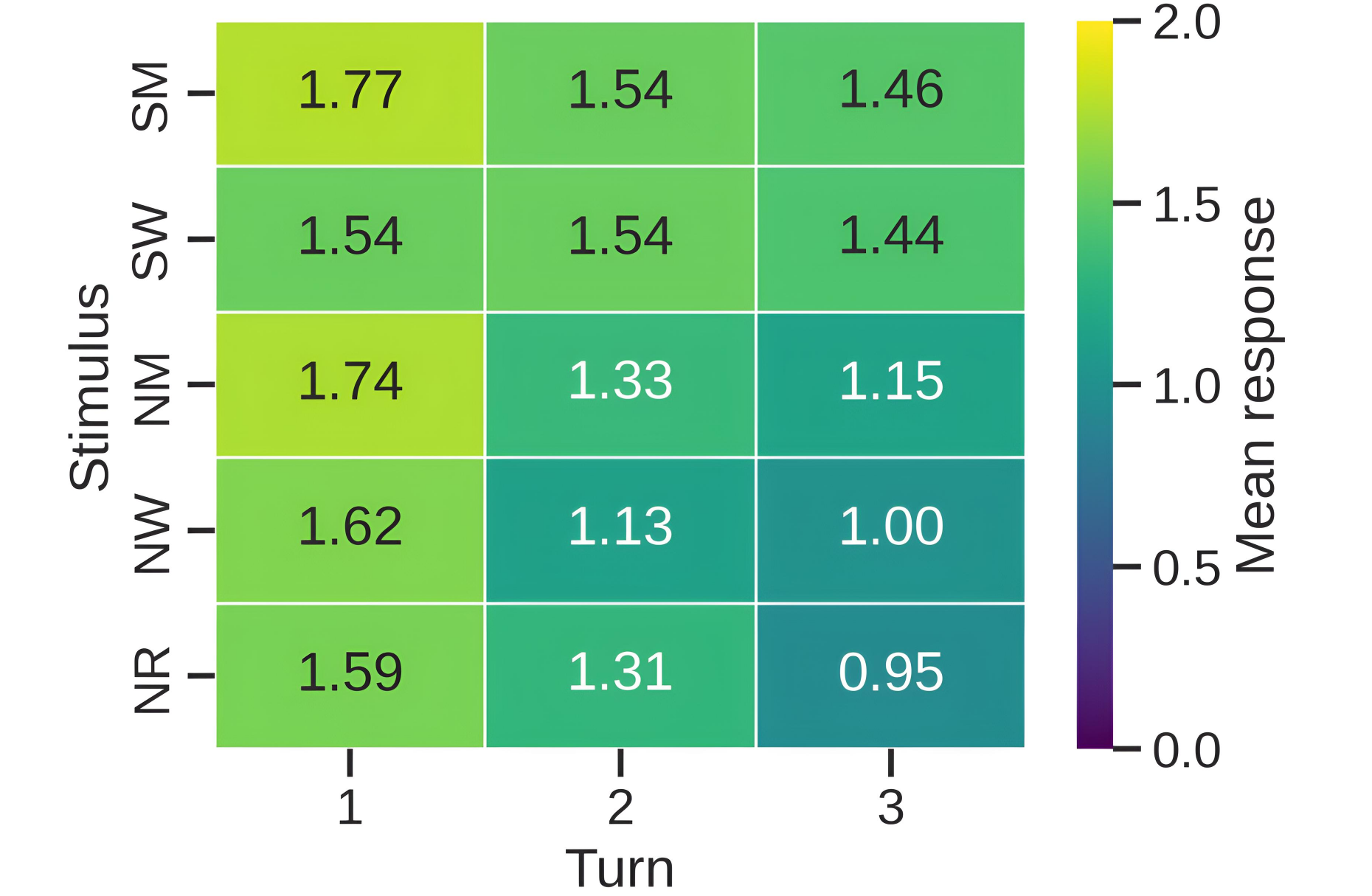}
\caption{Typical group}
\label{Fig_TurnMap_Typical_mean}
\end{subfigure}
\hfill
\begin{subfigure}[t]{0.48\textwidth}
\centering
\includegraphics[width=\linewidth]{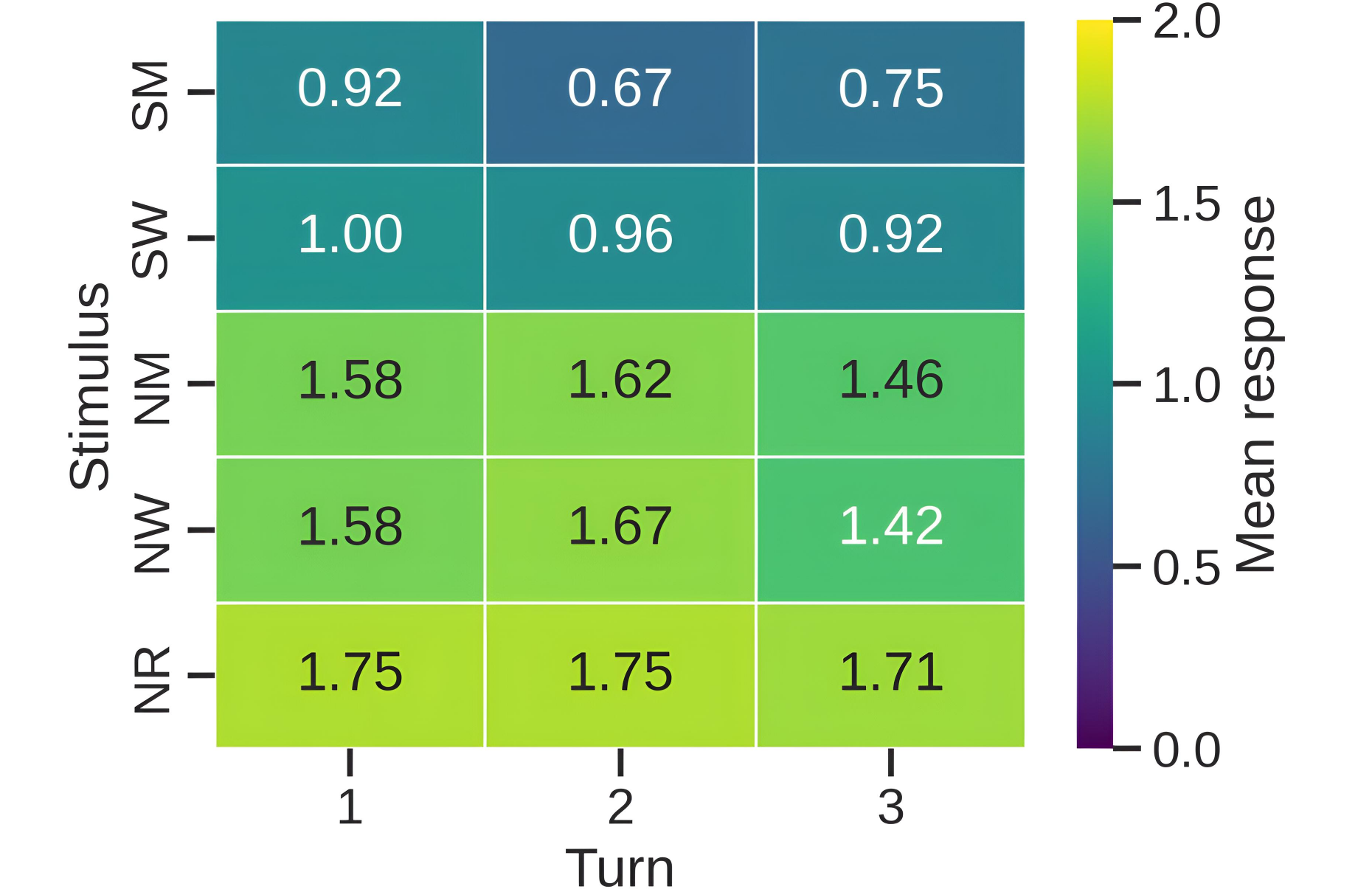}
\caption{ASD group}
\label{Fig_TurnMap_Autistic_mean}
\end{subfigure}

\caption{Turn-wise response maps across the three name-calling trials. 
(a) Typical group responses. 
(b) ASD group responses. }

\label{Fig_TurnMap}
\end{figure*}

\begin{figure}[!t]
    \centering
    \includegraphics[width=0.9\linewidth]{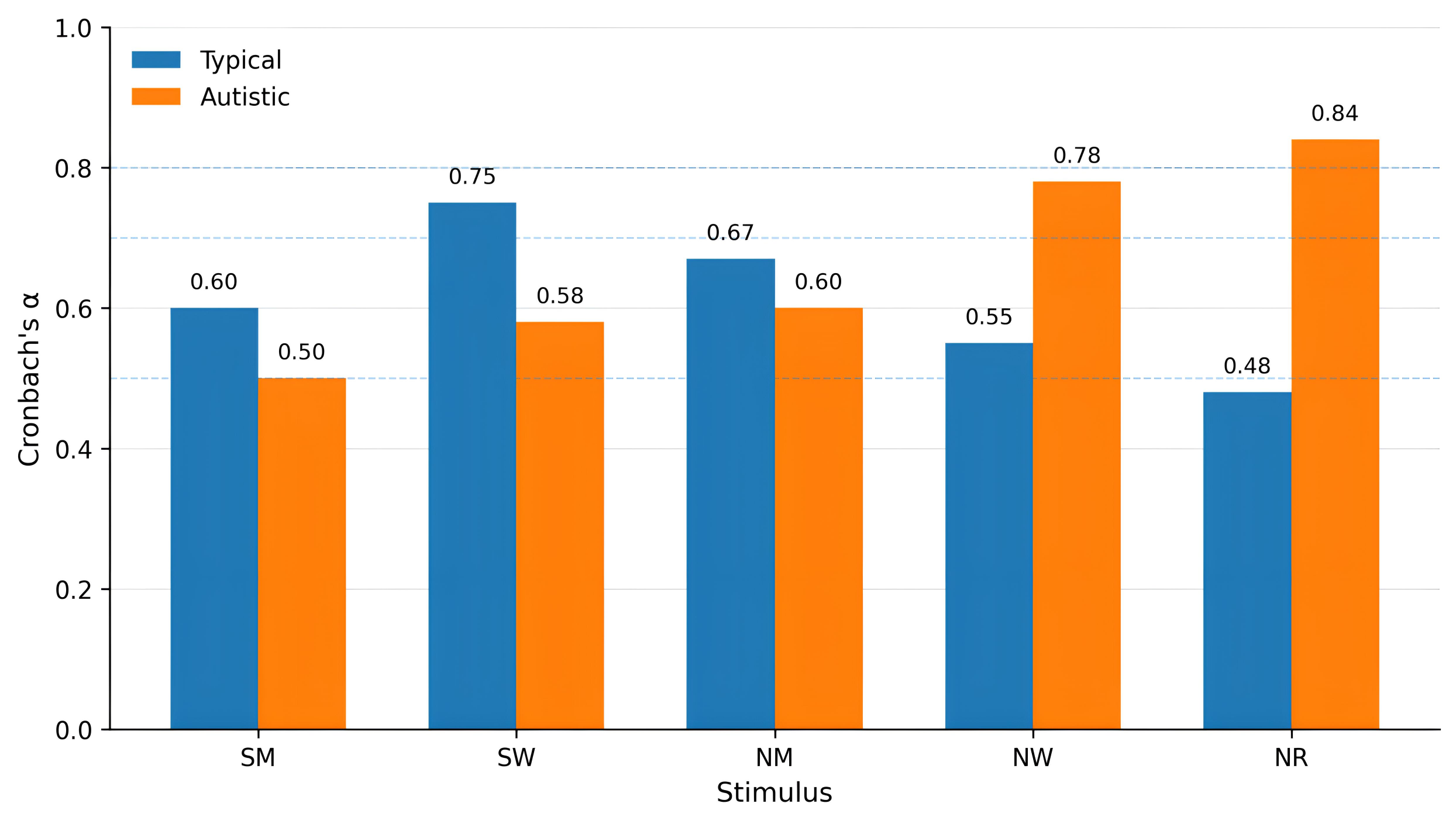}
    \caption{Cronbach’s $\alpha$ values across stimulus conditions for the typical and autistic groups.}
    \label{fig:alpha_reliability}
\end{figure}

These patterns are visually supported by the turn-wise response heatmaps in Fig.~\ref{Fig_TurnMap}, where preferred stimuli exhibit relatively stable response magnitudes across turns. The corresponding reliability values are summarized in Fig.~\ref{fig:alpha_reliability}.

\subsubsection{Signal Detection and Integrated Behavioral Summary}

Signal-detection sensitivity ($d'$) quantified group separability, with the highest values observed for NR ($d' = 1.46$) and NW ($d' = 1.31$), moderate sensitivity for NM ($d' = 0.98$), and comparatively lower values for SM and SW ($d' \approx 0.4$).

Integrating all analytical dimensions, TD children showed stronger but transient responses to human voices, whereas autistic children exhibited more sustained responses to robotic voices. Table~\ref{tab:summary_metrics} summarizes the dominant patterns across metrics.

\begin{table}[h!]
\centering
\caption{Summary of group-dominant metrics across analytical dimensions.}
\begin{tabular}{p{1.5cm}p{1.5cm}p{1.5cm}p{2cm}}
\hline
\textbf{Dimension} & \textbf{Dominant in TD} & \textbf{Dominant in ASD} & \textbf{Interpretation} \\
\hline
Response magnitude & Human voices (SM, SW) & Robot voices (NM, NW, NR) & Modality-specific sensitivity \\
Reliability ($\alpha$) & High for human & High for robot & Temporal consistency preference \\
Turn slope ($S$) & Negative & Flat or positive & Habituation vs sustained engagement \\
Variance & Low for human & Low for robot & Selective stability \\
Signal discrimination ($d'$) & Low & High for robot stimuli & Diagnostic separability \\
\hline
\end{tabular}
\label{tab:summary_metrics}
\end{table}

\subsection{Video-based analysis of NAO-related voices}
From the previous analysis, it is evident that autistic children show a preference for NAO robot-related voices; here, the video-based analysis of parameters NM, NW, and NR is continued for the two groups.

\subsubsection{Between-group Differences}
Non-parametric Mann--Whitney $U$ analyses (Table~\ref{tab:between}) revealed clear distinctions between autistic and typically developing (TD) participants in engagement-related metrics. 
The \textbf{Mean Eye-Openness Percentage (EOP)} scores were significantly higher in the autistic group across all three robotic stimuli: 
NM ($U = 1210.5$, $p < 1 \times 10^{-9}$, $d = -1.08$), 
NW ($U = 1211.0$, $p < 1 \times 10^{-9}$, $d = -1.11$), 
and NR ($U = 795.0$, $p < 1 \times 10^{-14}$, $d = -1.55$). 
Neither \textbf{Latency} nor \textbf{Duration} showed significant group differences ($p > 0.16$). 

\begin{table}[!h]
\centering
\caption{Between-group Mann--Whitney $U$ analysis across stimuli.}
\label{tab:between}
\begin{tabular}{p{1cm}p{1cm}p{1cm}p{1.4cm}p{1.4cm}}
\hline
\textbf{Stimulus} & \textbf{Metric} & \textbf{$U$} & \textbf{$p$-value} & \textbf{Cohen's $d$} \\
\hline
NM & Mean EOP & 1210.5 & $<1\times10^{-9}$ & $-1.08$ \\
NW & Mean EOP & 1211.0 & $<1\times10^{-9}$ & $-1.11$ \\
NR & Mean EOP & 795.0 & $<1\times10^{-14}$ & $-1.55$ \\
NM, NW, NR & Latency, Duration & ns & $>0.16$ & -- \\
\hline
\end{tabular}
\end{table}
 
\begin{figure*}[!t]
\centering
\includegraphics[width=0.45\textwidth]{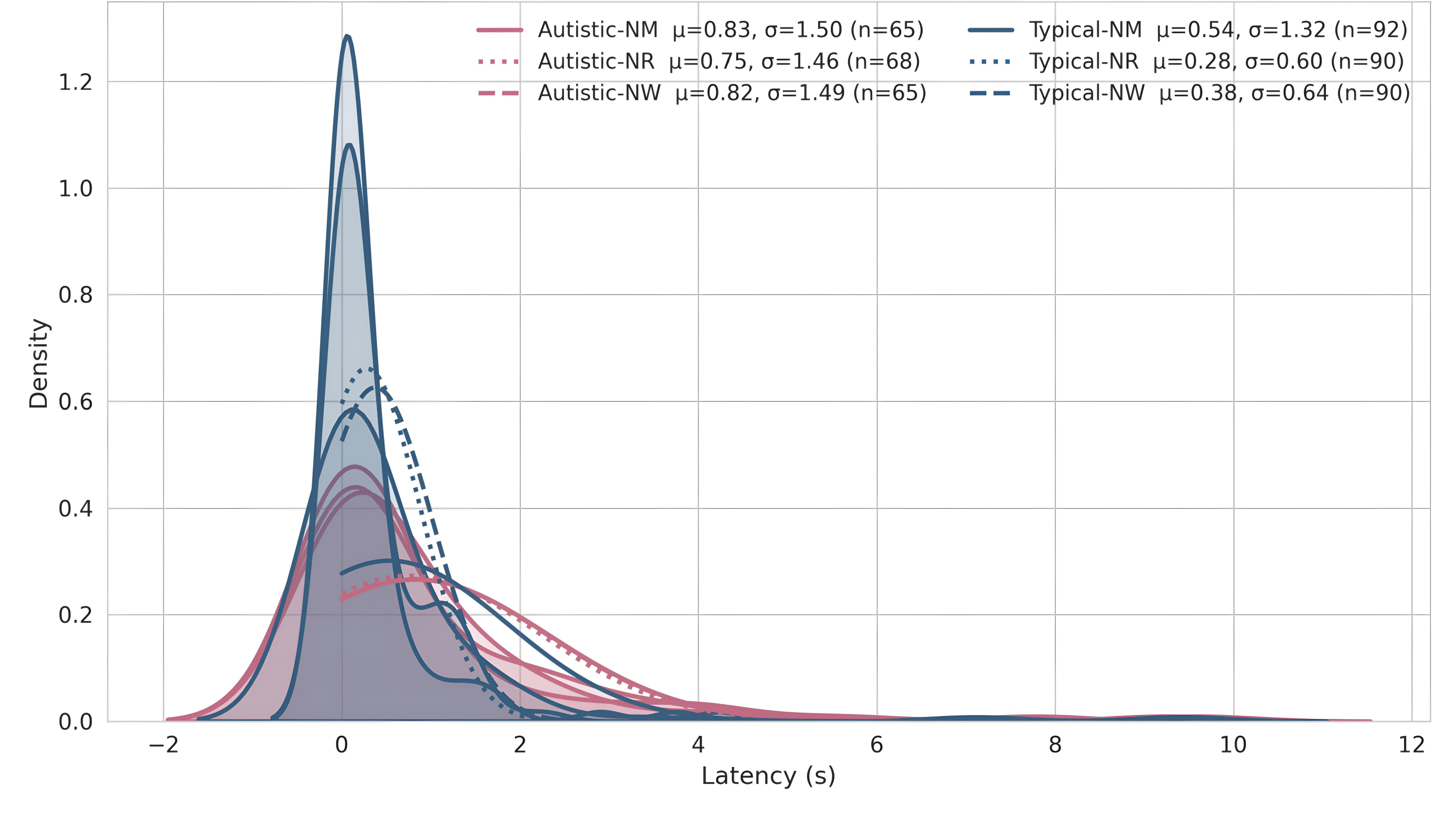}
\includegraphics[width=0.45\textwidth]{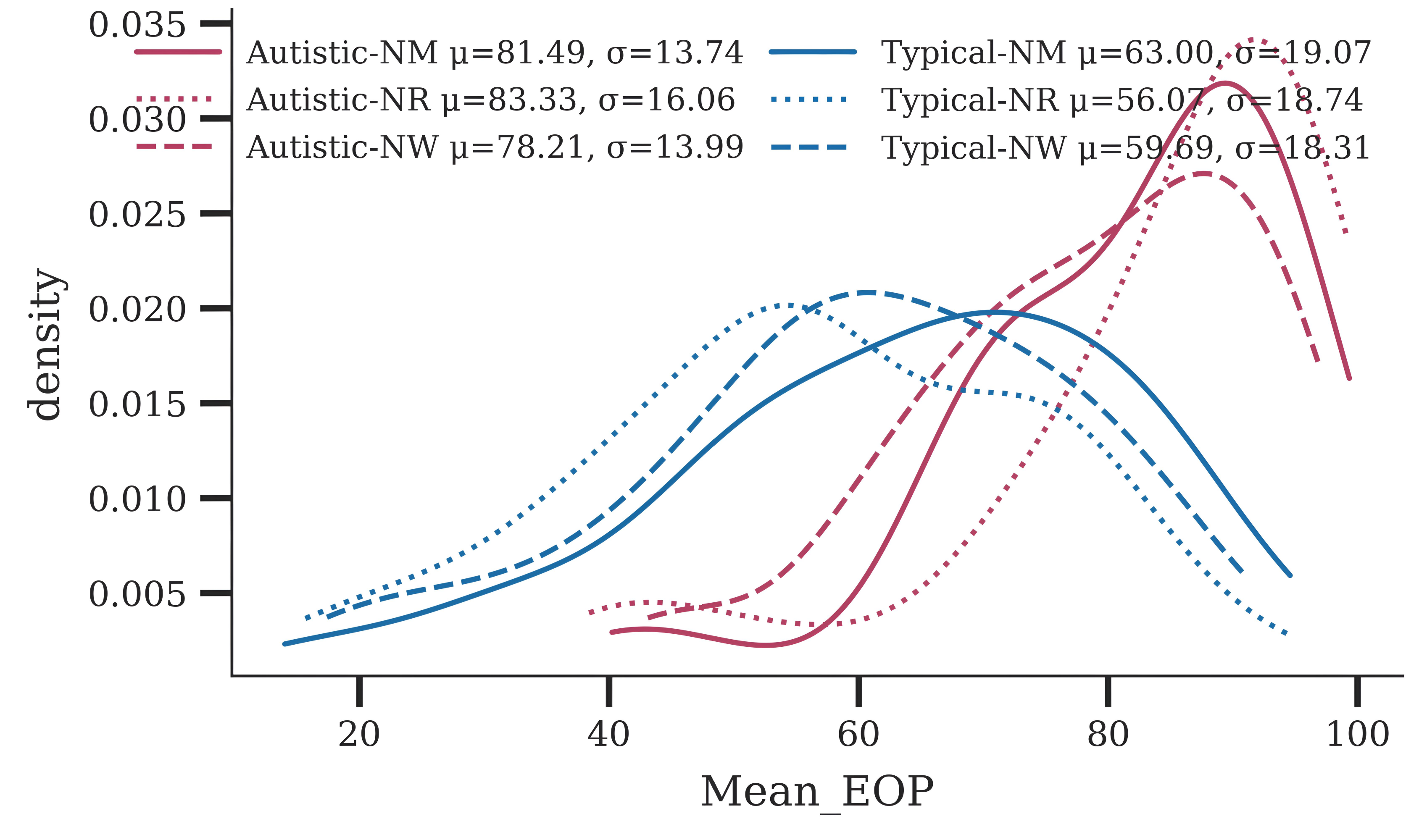}
\caption{(a) Latency and (b) Mean EOP distribution curves showing deviation from normality, justifying rank-based analyses.}
\label{fig:bellcurve_joint}
\end{figure*}

The bell curves (Fig.~\ref{fig:bellcurve_joint}) confirm the non-normal distribution patterns in Latency and Mean EOP, supporting the use of non-parametric tests.  

\subsubsection{Within-group Stimulus Effects}
\begin{figure}[!h]
\centering
\includegraphics[width=\columnwidth]{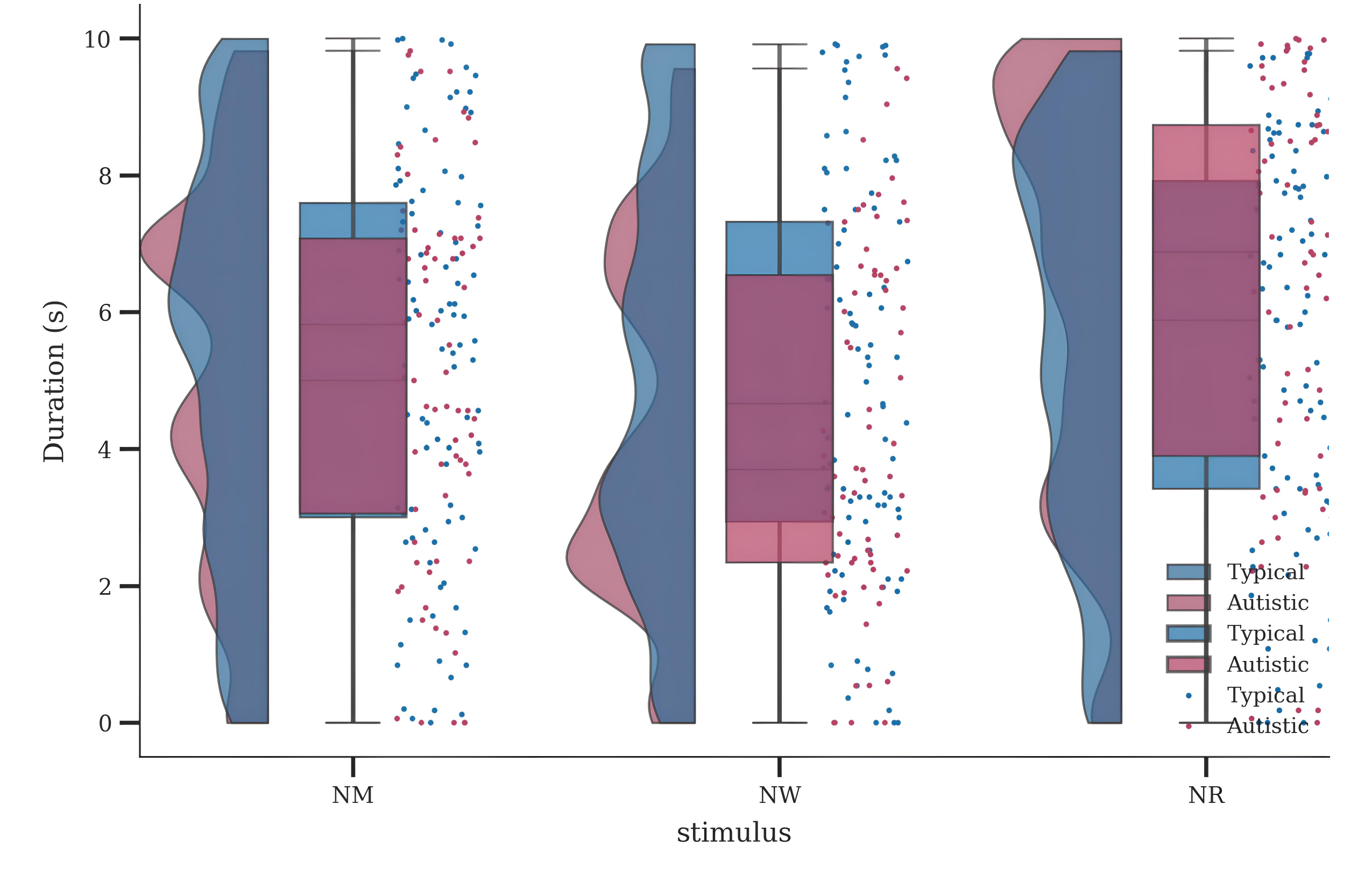}
\caption{Raincloud plots showing Duration distributions across stimuli for autistic and typical participants. Longer durations under NR are evident for autistic children.}
\label{fig:raincloud_duration}
\end{figure}

Kruskal--Wallis and Holm-adjusted post-hoc Mann--Whitney tests (Table~\ref{tab:within}) revealed stimulus-specific variations within each cohort. 
For autistic participants, significant effects were found for both \textbf{Duration} ($H = 19.56$, $p = 5.7 \times 10^{-5}$) and \textbf{Mean EOP} ($H = 8.74$, $p = 0.0126$). 
Post-hoc contrasts indicated that NR elicited longer interaction durations (NW vs.\ NR, $\Delta = -3.18$~s, $d = -0.75$) and higher engagement (NW vs.\ NR, $\Delta = -8.82$, $d = -0.34$). 
In contrast, TD children displayed a smaller yet significant effect for Mean EOP ($H = 6.80$, $p = 0.0334$), with higher values under NM than NR ($\Delta = +10.46$, $d = +0.37$). 
Latency remained non-significant across all comparisons ($p > 0.40$). 
The distribution of Duration across stimuli is shown in Figure~\ref{fig:raincloud_duration}, which highlights NR’s higher median values among autistic participants.

\begin{table}[h!]
\centering
\caption{Within-group Kruskal--Wallis and post-hoc comparisons.}
\label{tab:within}
\begin{tabular}{p{1cm}p{1cm}p{1cm}p{1.5cm}p{1.5cm}}
\hline
\textbf{Group} & \textbf{Metric} & \textbf{$H$} & \textbf{$p$-value} & \textbf{Key Contrast} \\
\hline
Autistic & Duration & 19.56 & $5.7\times10^{-5}$ & NR $>$ NW \\
Autistic & Mean EOP & 8.74 & 0.0126 & NR $>$ NW \\
Typical & Mean EOP & 6.80 & 0.0334 & NM $>$ NR \\
Both & Latency & -- & $>0.40$ & not significant \\
\hline
\end{tabular}
\end{table}

\subsubsection{Correlational Structure}
Spearman rank correlations (Table~\ref{tab:correlation}) revealed consistent relationships among behavioral indices in both groups. 
A significant negative association was observed between \textbf{Latency} and \textbf{Duration} (Autistic $\rho = -0.47$, $p = 4.8 \times 10^{-12}$; Typical $\rho = -0.45$, $p = 3.8 \times 10^{-15}$), indicating a speed–engagement tradeoff in which faster responses were typically followed by shorter interaction durations. 
In contrast, \textbf{Duration} and \textbf{Mean EOP} were positively correlated (Autistic $\rho = 0.28$, $p = 5.5 \times 10^{-5}$; Typical $\rho = 0.35$, $p = 1.9 \times 10^{-9}$), suggesting that longer interaction periods were associated with increased eye openness and sustained engagement. 
No significant association was observed between \textbf{Latency} and \textbf{Mean EOP} ($p > 0.6$), indicating that response speed and engagement intensity operated as largely independent behavioral dimensions.

\begin{table}[h!]
\centering
\caption{Spearman correlations among behavioral metrics.}
\label{tab:correlation}
\begin{tabular}{lccc}
\hline
\textbf{Pair} & \textbf{Group} & \textbf{$\rho$} & \textbf{$p$-value} \\
\hline
Latency--Duration & Autistic & $-0.47$ & $4.8\times10^{-12}$ \\
Latency--Duration & Typical & $-0.45$ & $3.8\times10^{-15}$ \\
Duration--EOP & Autistic & $0.28$ & $5.5\times10^{-5}$ \\
Duration--EOP & Typical & $0.35$ & $1.9\times10^{-9}$ \\
Latency--EOP & Both & ns & $>0.6$ \\
\hline
\end{tabular}
\end{table}

\subsubsection{Reliability and Distributional Checks}
Cronbach’s $\alpha$ coefficients (Table~\ref{tab:reliability_full}) assessed internal consistency of behavioral metrics across stimuli and turns. 
Both groups showed strong reliability for \textbf{Duration} (ASD $\alpha = 0.726$–$0.825$; TD $\alpha = 0.668$–$0.780$), indicating stable patterns of interaction persistence once engagement was initiated. 
Reliability for \textbf{Mean EOP} was moderate to high, with consistently higher values in the autistic group ($\alpha = 0.669$–$0.836$) compared to the typical group ($\alpha = 0.395$–$0.717$), suggesting more uniform engagement intensity among autistic participants. 
In contrast, \textbf{Latency} exhibited substantial variability, particularly in the typical group ($\alpha = 0.146$–$0.361$), indicating less consistent initiation timing. 
However, latency reliability increased for autistic participants under robotic voice conditions (NR; $\alpha \approx 0.81$), suggesting that predictable robotic stimuli produced more consistent response initiation in the autistic group. 
Overall, the findings suggest that engagement persistence was robust across groups, whereas initiation dynamics exhibited greater stimulus sensitivity, particularly within the autistic group. 

\begin{table*}[h!]
\centering
\caption{Reliability coefficients (Cronbach’s $\alpha$) for behavioral metrics across groups.}
\label{tab:reliability_full}
\begin{tabular}{p{3cm}p{3.5cm}p{3.5cm}}
\hline
\textbf{Variable} & \textbf{Autistic (range)} & \textbf{Typical (range)} \\
\hline
Duration & 0.726--0.825 & 0.668--0.780 \\
Mean EOP & 0.669--0.836 & 0.395--0.717 \\
Latency & 0.224--0.806 & 0.146--0.361 \\
\hline
\end{tabular}
\end{table*}

\subsubsection{Temporal Dynamics Across Turns}
Temporal adaptation was analyzed using per-participant slopes ($\beta_1$) for each metric and stimulus across turns (Table~\ref{tab:trend_full}). 
Among typical children, \textbf{Duration} exhibited strong negative slopes under NM ($-0.94$) and NW ($-1.15$), indicating habituation, and a smaller decline under NR ($-0.25$). 
Their \textbf{Latency} slopes were slightly positive ($+0.01$ to $+0.10$), suggesting marginally faster responses across turns. 
\textbf{Mean EOP} showed the sharpest decline for NM ($-3.13$) and NW ($-7.17$), but increased under NR ($+2.63$), reflecting recovery of engagement toward the robotic voice.  

Autistic participants showed shallower slopes overall.  
For \textbf{Duration}, NM and NW decreased mildly ($-0.14$, $-0.40$) but stabilized under NR ($+0.02$).  
\textbf{Latency} remained nearly flat (within $\pm0.01$), indicating minimal learning or fatigue trend.  
\textbf{Mean EOP} declined for NM ($-0.07$) and NW ($-0.90$) but increased for NR ($+0.56$).  
Together, these results indicate that while TD children exhibit clear habituation to human-like voices, autistic children maintain or slightly enhance engagement with robotic stimuli, especially with the NAO robotic voice.  

The full overview of analytical models and corresponding findings is presented in Table ~\ref{tab:overview}.

\begin{table*}[t]
\centering
\caption{Temporal slopes ($\beta_1$) across stimuli and groups with interpretation.}
\label{tab:trend_full}
\begin{tabular}{p{2cm}p{2cm}p{2cm}p{2cm}p{2cm}p{4cm}}
\hline
\textbf{Group} & \textbf{Variable} & \textbf{Stimulus NM} & \textbf{Stimulus NW} & \textbf{Stimulus NR} & \textbf{Interpretation} \\
\hline
Autistic & Duration & $-0.14$ & $-0.40$ & $+0.02$ & Duration slightly decreases for NM/NW but stabilizes under NR. \\
Autistic & Latency & $-0.01$ & $+0.01$ & $-0.01$ & Essentially flat; almost no learning trend.\\
Autistic & Mean EOP & $-0.07$ & $-0.90$ & $+0.56$ & Engagement declines slightly under NM,NW, but increases under NR.\\
Typical & Duration & $-0.94$ & $-1.15$ & $-0.25$ & Strong habituation; Drop with turn \\
Typical & Latency & $+0.01$ & $+0.10$ & $+0.01$ & Slight positive slopes; marginally faster response initiation. \\
Typical & Mean EOP & $-3.13$ & $-7.17$ & $+2.63$ & Clear divergence: engagement decays under NM/NW, rises under NR. \\
\hline
\end{tabular}
\end{table*}

\begin{table*}[h!]
\centering
\caption{Overview of analytical models and corresponding findings.}
\label{tab:overview}
\begin{tabular}{p{3cm} p{4cm} p{4cm} p{3cm}}
\hline
\textbf{Analytical Level} & \textbf{Model/Test Applied} & \textbf{Key Outcome} & \textbf{Supporting Output} \\
\hline
Between-group & Mann--Whitney $U$ & Autistic $>$ Typical for Mean EOP across all stimuli & Table~\ref{tab:between}, Fig.~\ref{fig:bellcurve_joint} \\
Within-group & Kruskal--Wallis + Holm MWU & NR $>$ NW for Autistic, NM $>$ NR for Typical & Table~\ref{tab:within}, Fig.~\ref{fig:raincloud_duration} \\
Correlation & Spearman $\rho$ & Negative (Latency–Duration), Positive (Duration–EOP) & Table~\ref{tab:correlation} \\
Reliability & Cronbach $\alpha$ & High for Duration; moderate for EOP; variable for Latency & Table~\ref{tab:reliability_full} \\
Temporal trends & Linear slope $\beta_1$ & Sustained engagement under NR; habituation under NM/NW & Table~\ref{tab:trend_full} \\
\hline
\end{tabular}
\end{table*}

\section{Conclusion}

The present study examined how stimulus modality, voice gender, and acoustic patterning influence name-calling responses among typically developing (TD) children and children with ASD. By integrating \textbf{manual ordinal ratings} with \textbf{frame-wise computational video analysis}, the investigation captured both categorical and temporal aspects of engagement. The analytical sequence, beginning with intra-group benchmarking in TD and progressing to cross-group contrasts, enabled the isolation of social versus perceptual factors that govern attention and responsiveness. 

Among TD participants, a clear engagement gradient emerged across five stimuli: human male (SM), human female (SW), NAO male (NM), NAO female (NW), and NAO robotic (NR). Significant modality effects were observed for all indices: Mean Eye-Openness Percentage (EOP; $H = 132.87,\, p < 10^{-30}$), Duration ($H = 103.21,\, p < 2 \times 10^{-21}$), and Latency ($H = 35.47,\, p = 2.4 \times 10^{-7}$). Human voices, particularly SM ($68.42 \pm 9.14\%$ EOP; $5.21 \pm 1.37$ s duration), evoked the highest expressivity, while SW ($64.87 \pm 10.02\%$) showed similar yet slightly lower engagement. This gender asymmetry within the human domain suggests greater orienting toward lower-pitched or assertive male prosody, consistent with reports that male voices elicit stronger attentional gating in early childhood~\cite{doi:10.3233/RNN-2010-0499}. 

In contrast, robotic voices elicited lower EOP but longer durations, especially NR ($44.65 \pm 7.11\%$ EOP; $6.81 \pm 1.68$ s duration), indicating rapid orientation coupled with sustained, low-arousal vigilance. Thus, TD engagement is characterized by \textit{affective intensity under human prosody} and \textit{attentional persistence under robotic regularity}.

Cross-group analyses revealed a pronounced \textbf{modality inversion}. TD children responded maximally to human voices (SM = $1.59 \pm 0.66$; SW = $1.50 \pm 0.82$; full response $\sim 70\%$), whereas ASD participants exhibited the highest responsiveness to robotic voices, particularly NR ($1.74 \pm 0.56$; $79\%$ full response). Cronbach’s $\alpha$ confirmed this dissociation (TD: $\alpha = 0.75$ for SW; ASD: $\alpha = 0.84$ for NR). These findings empirically support dual motivational pathways: \textit{social-reward-driven engagement} in TD versus \textit{predictability-driven stabilization} in ASD, aligning with the Social Motivation Theory~\cite{Chevallier2012SocialMotivationAutism} and predictive-coding accounts of autism~\cite{10.3389/fnhum.2014.00302}.

Gender-specific differentiation further illuminated these mechanisms. TD participants exhibited stronger and more reliable responses to male voices when both full and partial responses were considered alongside video-based analysis, whereas ASD participants displayed negligible differences between NAO’s male and female variants, implying that once affective ambiguity is minimized, \textit{voice gender ceases to modulate attention}.

Manual coding further indicated that the maximum response was $23.08\%$ for NR in TD participants and $51.39\%$ for SM in ASD participants. Collectively, these observations yield a \textbf{cue-complexity hypothesis}: TD engagement scales positively with social and prosodic richness, whereas ASD engagement scales inversely with it. 

Another observation was that, among TD participants, responses to the NM stimulus were higher than those to the NW stimulus. This difference was not comparably pronounced among ASD participants. One possible interpretation is that TD children were more likely to attribute a gender identity to the NAO robot and perceived its appearance and voice as more closely resembling a male figure. Informal feedback collected after the trials indicated that several TD participants reported perceiving the robot as ``male-like.'' In contrast, ASD participants appeared less likely to assign such gendered social attributes to the robot, which may explain the absence of a similarly strong NM--NW response difference in the ASD group.

Video-based robotic-only analyses (NM, NW, NR) reinforced these trends. Autistic participants displayed significantly higher Mean EOP across all robotic voices (NM: $U = 1210.5,\, p < 10^{-9}$; NW: $U = 1211.0,\, p < 10^{-9}$; NR: $U = 795.0,\, p < 10^{-14}$). Within-group contrasts showed NR elicited the longest durations and highest EOP in ASD ($\Delta \mathrm{EOP} = +8.8\%$, $\Delta \mathrm{Dur} = +3.2$ s, $p < 0.01$), while TD children favored NM over NR ($\Delta \mathrm{EOP} = +10.5$, $p = 0.0334$). 

Turn-wise slopes captured distinct temporal adaptations: TD responses declined across repetitions ($S_{\mathrm{TD}} = -0.22$), indicating habituation, whereas ASD responses stabilized or increased for NR ($S_{\mathrm{ASD}} = +0.08$), denoting sustained attentional entrainment under predictable acoustic cues~\cite{Klin2009BiologicalMotionAutism, doi:10.1073/pnas.1416797111}. The decline in TD responses was more pronounced for NAO-related stimuli, while responses to SM and SW remained relatively stable, likely reflecting adherence to social norms associated with responding to human callers. This effect appeared attenuated in ASD participants.

The computational pipeline ensured objective quantification and reproducibility. Frame-level analyses used \texttt{OpenCV}, \texttt{MediaPipe}, \texttt{MTCNN}, and \texttt{Dlib} for facial landmark extraction (468 points), from which EOP, latency, and duration were derived. Statistical evaluations employed Kruskal--Wallis and Mann--Whitney $U$ tests with Holm corrections, supplemented by Spearman correlations and effect-size estimates (Cohen’s $d$, $d'$). Visualization was implemented using \texttt{Matplotlib}, \texttt{Seaborn}, and \texttt{JoyPy}. Reliability ($\alpha = 0.67$--$0.84$) confirmed temporal stability, while signal-detection metrics ($d' = 1.46$ for NR) established robotic name calls as the most diagnostic discriminator between groups. This methodological synthesis bridges manual behavioral coding with algorithmic perception, contributing to reproducible computational frameworks in social robotics.

Integrating these dimensions, TD children showed \textit{strong but transient} affective engagement with socially rich human voices—particularly male prosody—whereas autistic children demonstrated \textit{stable and sustained} attention toward predictable robotic tones, largely insensitive to voice gender. Variance, reliability, and correlational analyses consistently underscored selective internal stability: TD variance was lowest for human cues, and ASD variance was lowest for robotic cues. These findings reveal complementary neurocognitive mechanisms—\textbf{affective entrainment} in TD and \textbf{predictive stabilization} in ASD—each optimizing engagement under distinct cue complexities. This align with prior work demonstrating that affective and engagement responses in children with ASD are highly heterogeneous as seen in the variability for human stimuli, thereby motivating personalized and adaptive modeling approaches in robot-mediated interaction \cite{Rudovic2018}.

From an applied standpoint, the results substantiate NAO’s \textbf{robotic voice} as a robust socially assistive interface. Its temporal regularity, controlled pitch contour, and low affective load minimized habituation and enhanced attentional reliability among ASD participants. Conversely, expressive modulation remains crucial for sustaining natural reciprocity in TD interactions.

Beyond confirming these modality effects, the present study advances an \textbf{early-stage framework} for tailoring robot-mediated social engagement in young autistic children, demonstrating that systematic voice modulation alone can enhance attentional stability even without exaggerated affect or facial mimicry. This approach represents a \textbf{translational step} toward adaptive human--robot systems that leverage predictability as a bridge to social attention. By establishing an empirical, data-driven link between voice acoustics and engagement reliability, this work contributes foundational evidence supporting the strategic use of robots such as NAO in structured early-age ASD interventions. Furthermore, response to one’s name, as a core and early-developing social communicative behavior, remains a widely recognized observational indicator for distinguishing atypical social responsiveness in children with ASD.

\textbf{Future Directions and Discussions.} Building on these findings, future research can expand this framework toward multimodal integration, combining facial, postural, and auditory channels for deeper modeling of child–robot engagement dynamics. Real-time analysis through onboard computation and adaptive feedback algorithms can enable NAO to adjust voice modulation, speech rate, and gaze timing contingent on live EOP fluctuations. Incorporating Large Language Model (LLM)-based interaction modules could also facilitate contextual name-calling and conversational adaptation, creating individualized engagement trajectories. Furthermore, combining this video-based approach with physiological measures such as galvanic skin response (GSR) or heart-rate variability may yield richer models of social arousal and predictability in autism. These extensions will advance both theoretical understanding of social attention and practical applications of AI-driven, emotionally aware robotics in therapeutic and educational contexts.

\section{Acknowledgement} 
We gratefully acknowledge the support of the Department of Mechanical Engineering and the Psychology Laboratory, Department of Humanities and Social Sciences, IIT Kanpur, for providing essential infrastructural facilities and interdisciplinary guidance throughout this research. We also extend our sincere thanks to the Kendriya Vidyalaya IIT Kanpur Campus, Pushpa Khanna Memorial Centre, and Amrita Rehabilitation Centre , Kanpur, for their collaboration, logistical assistance, and sustained cooperation during participant recruitment and data collection. Our deepest appreciation is extended to the participating children and their parents for their trust and voluntary involvement. 

We further acknowledge the dedicated efforts of the research assistants and staff members who supported the experimental setup, video processing, and data annotation across partner sites and our laboratory. We also sincerely thank Mr. Rishabh Pandey, Mr. Tanuj Gupta, Mr. Subhadeep Sahana, Mr. Sourabh Pandey, Ms. Malik Arsala Nissar, Ms. Samreen Zahoor, Mr. Abhishek Kumar Singh and Mr. Aman Sharma for their assistance during the experimental trials. We are deeply appreciative of all individuals and institutions whose collaboration and encouragement made this interdisciplinary research possible.

\section*{Declarations}

\subsection*{Ethics Approval and Consent to Participate}

All procedures performed in this study involving human participants were conducted in accordance with institutional ethical standards, IIT Kanpur (Approval No.\ [IITK/IEC/2024-25/II/25]). Informed consent was obtained from the participants' legal guardians prior to participation on a detailed consent form after a detailed briefing about the experiment.

\subsection*{Consent for Publication}

Informed consent for publication was obtained from the participants’ legal guardians, allowing the use of anonymized participant data.

\subsection*{Data Availability}

Due to the sensitive nature of the data involving children, the datasets generated and analyzed during the current study are not publicly available. Data may be available from the corresponding author upon reasonable request and subject to institutional and ethical approvals.

\subsection*{Conflict of Interest}

The authors declare no conflict of interest.

\nocite{*}
\bibliography{references}

@article{10.1001/archpedi.161.4.378,
    author = {Nadig, Aparna S. and Ozonoff, Sally and Young, Gregory S. and Rozga, Agata and Sigman, Marian and Rogers, Sally J.},
    title = {A Prospective Study of Response to Name in Infants at Risk for Autism},
    journal = {Archives of Pediatrics \& Adolescent Medicine},
    volume = {161},
    number = {4},
    pages = {378-383},
    year = {2007},
    month = {04},
    issn = {1072-4710},
    url = {https://doi.org/10.1001/archpedi.161.4.378},
}

@article{Zwaigenbaum2017ResponseName,
  author       = {Zwaigenbaum, Lonnie},
  title        = {Response to Name May Enhance Autism Spectrum Disorder Screening},
  journal      = {The Journal of Pediatrics},
  year         = {2017},
  volume       = {188},
  pages        = {308--311},
  publisher    = {Elsevier},
  doi          = {10.1016/j.jpeds.2017.06.058},
  issn         = {0022-3476}
}

@article{Dawson1998SocialOrientingAutism,
  author    = {Dawson, G. and Meltzoff, A. N. and Osterling, J. and Rinaldi, J. and Brown, E.},
  title     = {Children with Autism Fail to Orient to Naturally Occurring Social Stimuli},
  journal   = {Journal of Autism and Developmental Disorders},
  year      = {1998},
  volume    = {28},
  number    = {6},
  pages     = {479--485},
  month     = {December},
  doi       = {10.1023/a:1026043926488},
  pmid      = {9932234},
  issn      = {0162-3257},
  publisher = {Springer}
}

@Article{bs14020131,
AUTHOR = {Dubois-Sage, Marion and Jacquet, Baptiste and Jamet, Frank and Baratgin, Jean},
TITLE = {People with Autism Spectrum Disorder Could Interact More Easily with a Robot than with a Human: Reasons and Limits},
JOURNAL = {Behavioral Sciences},
VOLUME = {14},
YEAR = {2024},
NUMBER = {2},
ARTICLE-NUMBER = {131},
PubMedID = {38392485},
ISSN = {2076-328X},
DOI = {10.3390/bs14020131}
}

@article{Robaczewski2021NAOReview,
  author    = {Robaczewski, Adam and Bouchard, Julie and Bouchard, Kevin and Gaboury, S{\'e}bastien},
  title     = {Socially Assistive Robots: The Specific Case of the NAO},
  journal   = {International Journal of Social Robotics},
  year      = {2021},
  volume    = {13},
  number    = {4},
  pages     = {795--831},
  month     = {July},
  doi       = {10.1007/s12369-020-00664-7},
  issn      = {1875-4805},
  publisher = {Springer}
}

@INPROCEEDINGS{6190580,
  author={Ismail, Luthffi and Shamsuddin, Syamimi and Yussof, Hanafiah and Hashim, Hafizan and Bahari, Saiful and Jaafar, Ahmed and Zahari, Ismarrubie},
  booktitle={2011 IEEE International Conference on Control System, Computing and Engineering}, 
  title={Face detection technique of Humanoid Robot NAO for application in robotic assistive therapy}, 
  year={2011},
  volume={},
  number={},
  pages={517-521},
  doi={10.1109/ICCSCE.2011.6190580}}

@INPROCEEDINGS{8657569,
  author={Moghadas, M. and Moradi, H.},
  booktitle={2018 6th RSI International Conference on Robotics and Mechatronics (IcRoM)}, 
  title={Analyzing Human-Robot Interaction Using Machine Vision for Autism screening}, 
  year={2018},
  volume={},
  number={},
  pages={572-576},
  doi={10.1109/ICRoM.2018.8657569}}

@article{Osterling1994EarlyRecognitionAutism,
  author    = {Osterling, Julie and Dawson, Geraldine},
  title     = {Early Recognition of Children with Autism: A Study of First Birthday Home Videotapes},
  journal   = {Journal of Autism and Developmental Disorders},
  year      = {1994},
  volume    = {24},
  number    = {3},
  pages     = {247--257},
  month     = {June},
  doi       = {10.1007/BF02172225},
  issn      = {1573-3432},
  publisher = {Springer}
}

@article{Baranek1999AutismInfancyVideo,
  author    = {Baranek, Grace T.},
  title     = {Autism During Infancy: A Retrospective Video Analysis of Sensory-Motor and Social Behaviors at 9--12 Months of Age},
  journal   = {Journal of Autism and Developmental Disorders},
  year      = {1999},
  volume    = {29},
  number    = {3},
  pages     = {213--224},
  month     = {June},
  doi       = {10.1023/a:1023080005650},
  pmid      = {10425584},
  issn      = {0162-3257},
  publisher = {Springer}
}

@article{Werner2000RecognitionASDInfancy,
  author    = {Werner, Emily and Dawson, Geraldine and Osterling, Julie and Dinno, Nuhad},
  title     = {Brief Report: Recognition of Autism Spectrum Disorder Before One Year of Age: A Retrospective Study Based on Home Videotapes},
  journal   = {Journal of Autism and Developmental Disorders},
  year      = {2000},
  volume    = {30},
  number    = {2},
  pages     = {157--162},
  month     = {April},
  doi       = {10.1023/A:1005463707029},
  url       = {https://doi.org/10.1023/A:1005463707029},
  issn      = {1573-3432},
  publisher = {Springer}
}

@article{MILLER2017141,
title = {Response to Name in Infants Developing Autism Spectrum Disorder: A Prospective Study},
journal = {The Journal of Pediatrics},
volume = {183},
pages = {141-146.e1},
year = {2017},
issn = {0022-3476},
doi = {https://doi.org/10.1016/j.jpeds.2016.12.071},
author = {Meghan Miller and Ana-Maria Iosif and Monique Hill and Gregory S. Young and A.J. Schwichtenberg and Sally Ozonoff},
}

@article{HEDGER2020376,
title = {Social orienting and social seeking behaviors in ASD. A meta analytic investigation},
journal = {Neuroscience \& Biobehavioral Reviews},
volume = {119},
pages = {376-395},
year = {2020},
issn = {0149-7634},
doi = {https://doi.org/10.1016/j.neubiorev.2020.10.003},
author = {Nicholas Hedger and Indu Dubey and Bhismadev Chakrabarti}
}

@article{Robins2005RobotTherapyAutism,
  author    = {Robins, Ben and Dautenhahn, Kerstin and te Boekhorst, R. and Billard, Aude},
  title     = {Robotic Assistants in Therapy and Education of Children with Autism: Can a Small Humanoid Robot Help Encourage Social Interaction Skills?},
  journal   = {Universal Access in the Information Society},
  year      = {2005},
  volume    = {4},
  number    = {2},
  pages     = {105--120},
  month     = {December},
  doi       = {10.1007/s10209-005-0116-3},
  issn      = {1615-5297},
  publisher = {Springer}
}

@INPROCEEDINGS{6194716,
  author={Shamsuddin, Syamimi and Yussof, Hanafiah and Ismail, Luthffi and Hanapiah, Fazah Akhtar and Mohamed, Salina and Piah, Hanizah Ali and Zahari, Nur Ismarrubie},
  booktitle={2012 IEEE 8th International Colloquium on Signal Processing and its Applications}, 
  title={Initial response of autistic children in human-robot interaction therapy with humanoid robot NAO}, 
  year={2012},
  volume={},
  number={},
  pages={188-193},
  doi={10.1109/CSPA.2012.6194716}}

@INPROCEEDINGS{7006084,
  author={Miskam, Mohd Azfar and Shamsuddin, Syamimi and Samat, Mohd Ridzuan Abdul and Yussof, Hanafiah and Ainudin, Husna Ahmad and Omar, Abdul Rahman},
  booktitle={2014 International Symposium on Micro-NanoMechatronics and Human Science (MHS)}, 
  title={Humanoid robot NAO as a teaching tool of emotion recognition for children with autism using the Android app}, 
  year={2014},
  volume={},
  number={},
  pages={1-5},
  doi={10.1109/MHS.2014.7006084}}

@INPROCEEDINGS{7925401,
  author={Ackovska, Nevena and Kirandziska, Vesna and Tanevska, Ana and Bozinovska, Liljana and Božinovski, Adrijan},
  booktitle={SoutheastCon 2017}, 
  title={Robot - assisted therapy for autistic children}, 
  year={2017},
  volume={},
  number={},
  pages={1-2},
  doi={10.1109/SECON.2017.7925401}}

@inproceedings{inproceedings,
author = {Alemi, Minoo and Meghdari, Ali and Basiri, Nasim and Taheri, Alireza},
year = {2015},
month = {10},
pages = {},
title = {The Effect of Applying Humanoid Robots as Teacher Assistants to Help Iranian Autistic Pupils Learn English as a Foreign Language},
isbn = {978-3-319-25553-8},
doi = {10.1007/978-3-319-25554-5_1}
}

@INPROCEEDINGS{7451785,
  author={Beer, Jenay M. and Boren, Michelle and Liles, Karina R.},
  booktitle={2016 11th ACM/IEEE International Conference on Human-Robot Interaction (HRI)}, 
  title={Robot assisted music therapy a case study with children diagnosed with autism}, 
  year={2016},
  volume={},
  number={},
  pages={419-420},
  doi={10.1109/HRI.2016.7451785}}

@inproceedings{10.1145/3029798.3038354,
author = {Suzuki, Ryo and Lee, Jaeryoung and Rudovic, Ognjen},
title = {NAO-Dance Therapy for Children with ASD},
year = {2017},
isbn = {9781450348850},
publisher = {Association for Computing Machinery},
address = {New York, NY, USA},
doi = {10.1145/3029798.3038354},
booktitle = {Proceedings of the Companion of the 2017 ACM/IEEE International Conference on Human-Robot Interaction},
pages = {295–296},
numpages = {2},
location = {Vienna, Austria},
series = {HRI '17}
}

@INPROCEEDINGS{7745218,
  author={Chevalier, Pauline and Martin, Jean-Claude and Isableu, Brice and Bazile, Christophe and Iacob, David-Octavian and Tapus, Adriana},
  booktitle={2016 25th IEEE International Symposium on Robot and Human Interactive Communication (RO-MAN)}, 
  title={Joint Attention using Human-Robot Interaction: Impact of sensory preferences of children with autism}, 
  year={2016},
  volume={},
  number={},
  pages={849-854},
  doi={10.1109/ROMAN.2016.7745218}}

@ARTICLE{9292923,
  author={Cao, Hoang-Long and Simut, Ramona Elena and Desmet, Naomi and De Beir, Albert and Van De Perre, Greet and Vanderborght, Bram and Vanderfaeillie, Johan},
  journal={IEEE Access}, 
  title={Robot-Assisted Joint Attention: A Comparative Study Between Children With Autism Spectrum Disorder and Typically Developing Children in Interaction With NAO}, 
  year={2020},
  volume={8},
  number={},
  pages={223325-223334},
  doi={10.1109/ACCESS.2020.3044483}}

@article{Barnes07022021,
author = {Jaclyn A. Barnes and Chung Hyuk Park and Ayanna Howard and Myounghoon Jeon},
title = {Child-Robot Interaction in a Musical Dance Game: An Exploratory Comparison Study between Typically Developing Children and Children with Autism},
journal = {International Journal of Human–Computer Interaction},
volume = {37},
number = {3},
pages = {249--266},
year = {2021},
publisher = {Taylor \& Francis},
doi = {10.1080/10447318.2020.1819667},
note ={PMID: 33767571},
}

@article{Ismail2019RoboticsAutismReview,
  author    = {Ismail, Luthffi Idzhar and Verhoeven, Thibault and Dambre, Joni and Wyffels, Francis},
  title     = {Leveraging Robotics Research for Children with Autism: A Review},
  journal   = {International Journal of Social Robotics},
  year      = {2019},
  volume    = {11},
  number    = {3},
  pages     = {389--410},
  month     = {June},
  doi       = {10.1007/s12369-018-0508-1},
  issn      = {1875-4805},
  publisher = {Springer}
}

@article{Saleh18082021,
author = {Mohammed A. Saleh and Fazah Akhtar Hanapiah and Habibah Hashim},
title = {Robot applications for autism: a comprehensive review},
journal = {Disability and Rehabilitation: Assistive Technology},
volume = {16},
number = {6},
pages = {580--602},
year = {2021},
publisher = {Taylor \& Francis},
doi = {10.1080/17483107.2019.1685016},
note ={PMID: 32706602},
URL = { https://doi.org/10.1080/17483107.2019.1685016},
}

@ARTICLE{9635826,
  author={Bartl-Pokorny, Katrin D. and Pykała, Małgorzata and Uluer, Pinar and Barkana, Duygun Erol and Baird, Alice and Kose, Hatice and Zorcec, Tatjana and Robins, Ben and Schuller, BJörn W. and Landowska, Agnieszka},
  journal={IEEE Access}, 
  title={Robot-Based Intervention for Children With Autism Spectrum Disorder: A Systematic Literature Review}, 
  year={2021},
  volume={9},
  number={},
  pages={165433-165450},
  doi={10.1109/ACCESS.2021.3132785}}

@article{ROMEROGARCIA2021103797,
title = {Q-CHAT-NAO: A robotic approach to autism screening in toddlers},
journal = {Journal of Biomedical Informatics},
volume = {118},
pages = {103797},
year = {2021},
issn = {1532-0464},
doi = {https://doi.org/10.1016/j.jbi.2021.103797},
author = {Rubén Romero-García and Rafael Martínez-Tomás and Pilar Pozo and Félix {de la Paz} and Encarnación Sarriá},
}

@INPROCEEDINGS{8373812,
  author={Baltrusaitis, Tadas and Zadeh, Amir and Lim, Yao Chong and Morency, Louis-Philippe},
  booktitle={2018 13th IEEE International Conference on Automatic Face \& Gesture Recognition (FG 2018)}, 
  title={OpenFace 2.0: Facial Behavior Analysis Toolkit}, 
  year={2018},
  volume={},
  number={},
  pages={59-66},
  doi={10.1109/FG.2018.00019}}

@Article{robotics7020025,
AUTHOR = {Di Nuovo, Alessandro and Conti, Daniela and Trubia, Grazia and Buono, Serafino and Di Nuovo, Santo},
TITLE = {Deep Learning Systems for Estimating Visual Attention in Robot-Assisted Therapy of Children with Autism and Intellectual Disability},
JOURNAL = {Robotics},
VOLUME = {7},
YEAR = {2018},
NUMBER = {2},
ARTICLE-NUMBER = {25},
ISSN = {2218-6581},
DOI = {10.3390/robotics7020025}
}

@article{https://doi.org/10.1002/aur.1527,
author = {Pennisi, Paola and Tonacci, Alessandro and Tartarisco, Gennaro and Billeci, Lucia and Ruta, Liliana and Gangemi, Sebastiano and Pioggia, Giovanni},
title = {Autism and social robotics: A systematic review},
journal = {Autism Research},
volume = {9},
number = {2},
pages = {165-183},
doi = {https://doi.org/10.1002/aur.1527},
year = {2016}
}

@article{doi:10.1126/scirobotics.aao6760,
author = {Ognjen Rudovic  and Jaeryoung Lee  and Miles Dai  and Björn Schuller  and Rosalind W. Picard },
title = {Personalized machine learning for robot perception of affect and engagement in autism therapy},
journal = {Science Robotics},
volume = {3},
number = {19},
pages = {eaao6760},
year = {2018},
doi = {10.1126/scirobotics.aao6760},
URL = {https://www.science.org/doi/abs/10.1126/scirobotics.aao6760},
}

@inproceedings{kozima2005interactive,
  title={Interactive robots for communication-care: A case-study in autism therapy},
  author={Kozima, Hideki and Nakagawa, Cocoro and Yasuda, Yuriko},
  booktitle={ROMAN 2005. IEEE International Workshop on Robot and Human Interactive Communication, 2005.},
  pages={341--346},
  year={2005},
  organization={IEEE}
}

@article{DBLP:journals/corr/abs-1906-08172,
  author       = {Camillo Lugaresi and
                  Jiuqiang Tang and
                  Hadon Nash and
                  Chris McClanahan and
                  Esha Uboweja and
                  Michael Hays and
                  Fan Zhang and
                  Chuo{-}Ling Chang and
                  Ming Guang Yong and
                  Juhyun Lee and
                  Wan{-}Teh Chang and
                  Wei Hua and
                  Manfred Georg and
                  Matthias Grundmann},
  title        = {MediaPipe: {A} Framework for Building Perception Pipelines},
  journal      = {CoRR},
  volume       = {abs/1906.08172},
  year         = {2019},
  eprinttype    = {arXiv},
  eprint       = {1906.08172},
  bibsource    = {dblp computer science bibliography, https://dblp.org}
}

@article{DBLP:journals/corr/BulatT17a,
  author       = {Adrian Bulat and
                  Georgios Tzimiropoulos},
  title        = {How far are we from solving the 2D {\\\&} 3D Face Alignment problem?
                  (and a dataset of 230, 000 3D facial landmarks)},
  journal      = {CoRR},
  volume       = {abs/1703.07332},
  year         = {2017},
  eprinttype    = {arXiv},
  eprint       = {1703.07332},
  bibsource    = {dblp computer science bibliography, https://dblp.org}
}

@article{cech2016real,
  title={Real-time eye blink detection using facial landmarks},
  author={Cech, Jan and Soukupova, Tereza},
  journal={Cent. Mach. Perception, Dep. Cybern. Fac. Electr. Eng. Czech Tech. Univ. Prague},
  pages={1--8},
  year={2016}
}

@article{Chevallier2012SocialMotivationAutism,
  author    = {Coralie Chevallier and Gregor Kohls and Vanessa Troiani and Edward S. Brodkin and Robert T. Schultz},
  title     = {The social motivation theory of autism},
  journal   = {Trends in Cognitive Sciences},
  year      = {2012},
  volume    = {16},
  number    = {4},
  pages     = {231--239},
  doi       = {10.1016/j.tics.2012.02.007},
  pmid      = {22425667},
  publisher = {Elsevier}
}

@ARTICLE{10.3389/fnhum.2014.00302,
    
AUTHOR={Lawson, Rebecca P.  and Rees, Geraint  and Friston, Karl J. },
           
TITLE={An aberrant precision account of autism},         
JOURNAL={Frontiers in Human Neuroscience},          
VOLUME={Volume 8 - 2014},  
YEAR={2014},
DOI={10.3389/fnhum.2014.00302},  
ISSN={1662-5161},
}

@article{doi:10.3233/RNN-2010-0499,
author = {Tobias Grossmann},
title ={The development of emotion perception in face and voice during
			 infancy},

journal = {Restorative Neurology and Neuroscience},
volume = {28},
number = {2},
pages = {219-236},
year = {2010},
doi = {10.3233/RNN-2010-0499},
    note ={PMID: 20404410},

URL = { https://doi.org/10.3233/RNN-2010-0499},


}

@article{
doi:10.1073/pnas.1416797111,
author = {Pawan Sinha  and Margaret M. Kjelgaard  and Tapan K. Gandhi  and Kleovoulos Tsourides  and Annie L. Cardinaux  and Dimitrios Pantazis  and Sidney P. Diamond  and Richard M. Held },
title = {Autism as a disorder of prediction},
journal = {Proceedings of the National Academy of Sciences},
volume = {111},
number = {42},
pages = {15220-15225},
year = {2014},
doi = {10.1073/pnas.1416797111},
URL = {https://www.pnas.org/doi/abs/10.1073/pnas.1416797111},
}

@article{Klin2009BiologicalMotionAutism,
  author    = {Ami Klin and David J. Lin and Phillip Gorrindo and Gordon Ramsay and Warren Jones},
  title     = {Two-year-olds with autism orient to non-social contingencies rather than biological motion},
  journal   = {Nature},
  year      = {2009},
  volume    = {459},
  number    = {7244},
  pages     = {257--261},
  doi       = {10.1038/nature07868},
  pmid      = {19329996},
  publisher = {Nature Publishing Group}
}

@article{Rudovic2018,

author = {Ognjen Rudovic  and Jaeryoung Lee  and Miles Dai  and Björn Schuller  and Rosalind W. Picard },
title = {Personalized machine learning for robot perception of affect and engagement in autism therapy},
journal = {Science Robotics},
volume = {3},
number = {19},
pages = {eaao6760},
year = {2018},
doi = {10.1126/scirobotics.aao6760},
}
\end{document}